\documentclass[12pt,preprint]{aastex6}
\usepackage[latin9]{inputenc}
\usepackage{color}
\usepackage{float}
\usepackage{textcomp}
\usepackage{amsmath}
\usepackage{graphicx}
\usepackage{setspace}
\usepackage{hyperref}
\usepackage{xcolor}

\shorttitle{The effect of rotation on fingering convection in stellar interiors} \shortauthors{Sengupta and Garaud}
\begin{document}
\title{The effect of rotation on fingering convection in stellar interiors}
\author{S.Sengupta and P. Garaud}
\affil{Department of Applied Mathematics and Statistics, Baskin School of Engineering, University of California Santa Cruz, 1156 High Street, Santa Cruz CA 95064} 
\begin{abstract}
We study the effects of rotation on the growth and saturation of the
double-diffusive fingering (thermohaline) instability at low Prandtl
number. Using direct numerical simulations, we estimate the compositional
transport rates as a function of the relevant non-dimensional parameters
- the Rossby number, inversely proportional to the rotation rate,
and the density ratio which measures the relative thermal and compositional
stratifications. Within our explored range of parameters, we generally
find rotation to have little effect on vertical transport. However,
we also present one exceptional case where a cyclonic large scale
vortex (LSV) is observed at low density ratio and fairly low Rossby
number. The LSV leads to significant enhancement in the fingering
transport rates by concentrating compositionally dense downflows at
its core. We argue that the formation of such LSVs could be relevant
to solving the missing mixing problem in RGB stars.
\end{abstract}

\section{Introduction\label{sec:into}}

Over the past decade or so, there has been a resurgence in interest
about the role of fingering convection as a mechanism for transport
of chemical species in the radiative zones of a variety of objects,
ranging from accreting main-sequence stars and white dwarfs in binary
systems (\cite{1997MNRAS.290..283M,1998MNRAS.301..699M}, \cite{2010Ap&SS.328..209T,2012ApJ...753...49V};
\cite{2007A&A...464L..57S}; \cite{2013ApJ...762....8D,2013A&A...557L..12D})
to exoplanet host stars (\cite{2004ApJ...605..874V,2011ApJ...728L..30G,2012ApJ...753...49V}),
as well as in the interiors of more evolved low-mass red-giant branch
(RGB) stars (\cite{2007A&A...467L..15C,2008ApJ...684..626D,2014A&A...570A..58W})
and possibly also in planetary atmospheres due to chemical reactions
(\cite{2015ApJ...804L..17T,2016ApJ...817L..19T}; although see \cite{2018ApJ...853L..30L}).
Recent numerical simulations of fingering convection by \cite{2010ApJ...723..563D,2011ApJ...727L...8D,2011JFM...677..530T,2013ApJ...768...34B}
(see the review by \cite{2018AnRFM..50..275G}) have consistently
shown that the typical values of mixing rates are two orders of magnitude
below those required to match observed abundance patterns in RGB stars
above the so-called ``luminosity bump'' (\cite{2000A&A...354..169G,2007A&A...467L..15C}).
The only way to reconcile theory and observations is to invoke the
existence of some previously unaccounted for mechanism that could
somehow significantly enhance mixing by fingering convection in these
stars (see, e.g. \cite{2014ApJ...792L..30M}, or \cite{2015ApJ...808...89G}
for some first attempts at cracking the problem).

The obvious candidates for such mechanisms in stars are rotation,
shear and magnetic fields. While the latter two remain to be explored
to date, the effect of rotation on oscillatory double-diffusive convection
(ODDC) has recently been studied in \cite{2017ApJ...834...44M} using
direct numerical simulations (DNSs) with the PADDI code (\cite{2008GGG.....9.5003S},
\cite{2011ApJ...728L..29T}, \cite{2011JFM...677..554S}). In this
paper, we apply the framework of \cite{2017ApJ...834...44M} to the
fingering regime and attempt to quantify the effect of rotation on
the growth and development of fingering instabilities in parameter
regimes relevant for stars. We begin by presenting the model setup
(Section 2) followed by a linear stability analysis of the fingering
instability in presence of rotation (Section 3) before quantifying
its effect in stellar interiors (Section 4) with the help of DNSs
(Section 5). We conclude in Sections 6 and 7 by discussing the relevance
of our findings for RGB stars. 

\section{The Model\label{sec:The-Model}}

In this work, we use the Boussinesq approximation (\cite{boussinesq1903theorie,1960ApJ...131..442S})
in a Cartesian setup which assumes constant background temperature
and composition gradients over the height of the computational domain,
and a linearized equation of state given by
\begin{equation}
\frac{\tilde{\rho}}{\rho_{0}}=-\alpha\tilde{T}+\beta\tilde{\mu},
\end{equation}
where $\tilde{\rho}$, $\tilde{T}$ and $\tilde{\mu}$ are the perturbations
to the background density, temperature and composition respectively
and $\rho_{0}$ is the mean density of the fluid in the region considered.
The coefficients $\alpha$ and $\beta$ are defined as 
\begin{alignat}{1}
\alpha & =-\frac{1}{\rho_{0}}\left.\frac{\partial\rho}{\partial T}\right|{}_{p,\mu},\,\beta=\frac{1}{\rho_{0}}\left.\frac{\partial\rho}{\partial\mu}\right|{}_{p,T},
\end{alignat}
where $p$ denotes pressure. We assume a constant background rotation
defined by the angular velocity vector $\mathbf{\Omega}=\varOmega\boldsymbol{e}_{\Omega}$,
with $\boldsymbol{e}_{\Omega}$ being the unit vector in the direction
of $\mathbf{\Omega}$ :
\begin{equation}
\boldsymbol{e}_{\Omega}=(0,\,\sin\theta,\,\cos\theta),\label{eq:omegaunitvector}
\end{equation}
where $\theta$ is the angle between the rotation axis and the \textit{z}-axis
of our domain, which is aligned with gravity. \\
Following \cite{2011ApJ...728L..29T}, we use the following units
for length $[l]$, time $[t]$, temperature $[T]$ and chemical composition
$[\mu]$:
\begin{gather*}
[l]=d=\left(\frac{\kappa_{T}\nu}{\alpha g|T_{0z}-T_{0z}^{ad}|}\right)^{\frac{1}{4}},\,[t]=\frac{d^{2}}{\kappa_{T}},\\
{}[T]=d|T_{0z}-T_{0z}^{ad}|,\,[\mu]=\frac{\alpha}{\beta}d|T_{0z}-T_{0z}^{ad}|,
\end{gather*}
where $g$ is the local acceleration due to gravity, $\nu$ is the
viscosity of the medium, $\kappa_{T}$ is the thermal diffusivity,
$T_{0z}$ is the background temperature gradient with respect to position
\textit{z} and $T_{0z}^{ad}=-\frac{g}{c_{p}}$ is the corresponding
adiabatic temperature gradient, where $c_{p}$ is the specific heat
at constant pressure. Using this choice of units, we can write the
non-dimensional form of the Navier-Stokes equations for the velocity
field $\mathbf{u}=(u,v,w)$ as follows:
\begin{alignat}{1}
\frac{1}{{\rm Pr}}\left[\left(\frac{\partial\mathbf{u}}{\partial t}+\mathbf{u}\cdot\nabla\mathbf{u}\right)+\sqrt{{\rm Ta}^{*}}\left(\boldsymbol{e}_{\Omega}\times\mathbf{u}\right)\right] & =-\nabla\widetilde{p}+(\widetilde{T}-\widetilde{\mu})\mathbf{e_{z}}+\nabla^{2}\mathbf{u},\label{eq:momeq}\\
\frac{\partial\widetilde{T}}{\partial t}+\mathbf{u}\cdot\nabla\widetilde{T}+w & =\nabla^{2}\widetilde{T},\label{eq:tempeq}\\
\frac{\partial\tilde{\mu}}{\partial t}+\mathbf{u}\cdot\nabla\widetilde{\mu}+\frac{w}{R_{0}} & =\tau\nabla^{2}\widetilde{\mu},\label{eq:goveqs}\\
\nabla\cdot\mathbf{u} & =0,\label{eq:incompressibility}
\end{alignat}
with four relevant non-dimensional parameters being the Prandtl number
(${\rm Pr}$), the diffusivity ratio ($\tau$), the density ratio
($R_{0}$) and the finger-based Taylor number ($\text{{\rm Ta}}^{*}$)
defined as (\cite{2017ApJ...834...44M}):
\begin{alignat}{1}
{\rm Pr} & =\frac{\nu}{\kappa_{T}},\,\tau=\frac{\kappa_{\mu}}{\kappa_{T}},\nonumber \\
R_{0} & =\frac{\alpha|T_{0z}-T_{0z}^{ad}|}{\beta\mu_{0z}},\nonumber \\
{\rm Ta}^{*} & =\frac{4\Omega^{2}d^{4}}{\kappa_{T}^{2}}.\label{eq:parameters}
\end{alignat}
As reviewed by \cite{2018AnRFM..50..275G}, the density ratio $R_{0}$
measures the effective stratification of the system, with $R_{0}=1$
corresponding to the limit of overturning convection. In non-rotating
stars, a region is unstable to basic fingering when
\begin{equation}
1<R_{0}<\frac{1}{\tau}.\label{eq:nonrotregime}
\end{equation}
The effect of rotation in turn is described by the finger-based Taylor
number ${\rm Ta^{*}}$ (see Section 4 for more detail on the significance
of ${\rm Ta^{*}}$ ). In what follows, we assume that the computational
domain is triply periodic, which greatly simplifies both the linear
stability analysis (Section 3) and the numerics (Section 5 and beyond).

\section{Linear stability analysis\label{sec:Linear-stability-analysis}}

We linearize the set of governing equations (Eq~\ref{eq:momeq} -
\ref{eq:incompressibility}) and use the ansatz:
\begin{equation}
q(x,y,z,t)=\hat{q}e^{i(lx+my+kz)+\lambda t},\label{eq:ansatz}
\end{equation}
for $q=\{\mathbf{u,\,}\widetilde{T},\,\widetilde{\mu}\}$. After some
algebra, we obtain a quartic polynomial equation for the growth rate
$\lambda$:
\begin{gather}
(\lambda+{\rm Pr}K^{2})^{2}(\lambda+\tau K^{2})(\lambda+K^{2})+\frac{k_{h}^{2}}{K^{2}}{\rm Pr}(\lambda+{\rm Pr}K^{2})[(\lambda+\tau K^{2})-R_{o}^{-1}(\lambda+K^{2})]\nonumber \\
+{\rm Ta}^{*}\frac{(m\sin\theta+k\cos\theta)^{2}}{K^{2}}(\lambda+\tau K^{2})(\lambda+K^{2})=0,\label{eq:dispersionrel}
\end{gather}
where $K=\sqrt{k_{h}^{2}+k^{2}}\mbox{ is the total wavenumber and }k_{h}=\sqrt{l^{2}+m^{2}}$
is the horizontal wavenumber. This is almost identical to the growth
rate equation obtained in the ODDC case (Eq~16 in \cite{2017ApJ...834...44M})
except for the sign in front of the second term (namely, the term
proportional to $\frac{k_{h}^{2}}{K^{2}}$) which is positive in the
fingering case, and negative in the ODDC case.

\subsection{Regime of Instability}

It can be shown that the fastest growing modes (i.e. modes with largest
$Re(\lambda)$ satisfying Eq~\ref{eq:dispersionrel}) have $k=0\mbox{ and }m=0$
(\cite{2017ApJ...834...44M}). Thus, these modes remain unaffected
by rotation (since the rotation term in Eq~\ref{eq:dispersionrel}
drops out for $k=m=0$), and the range of density ratios for which
fingering takes place is unchanged:
\begin{equation}
1<R_{0}<\frac{1}{\tau},
\end{equation}
where the lower limit of $R_{0}=1$ corresponds to the system being
unstable to overturning convection (Ledoux unstable) while the upper
limit $R_{0}=\frac{1}{\tau}$ corresponds to marginal stability to
fingering convection.

\subsection{Fastest growing modes\label{subsec:Fastest-growing-modes}}

\begin{figure}[H]
\raggedright{}%
\begin{minipage}[t]{0.45\columnwidth}%
\begin{flushleft}
\includegraphics[scale=0.33]{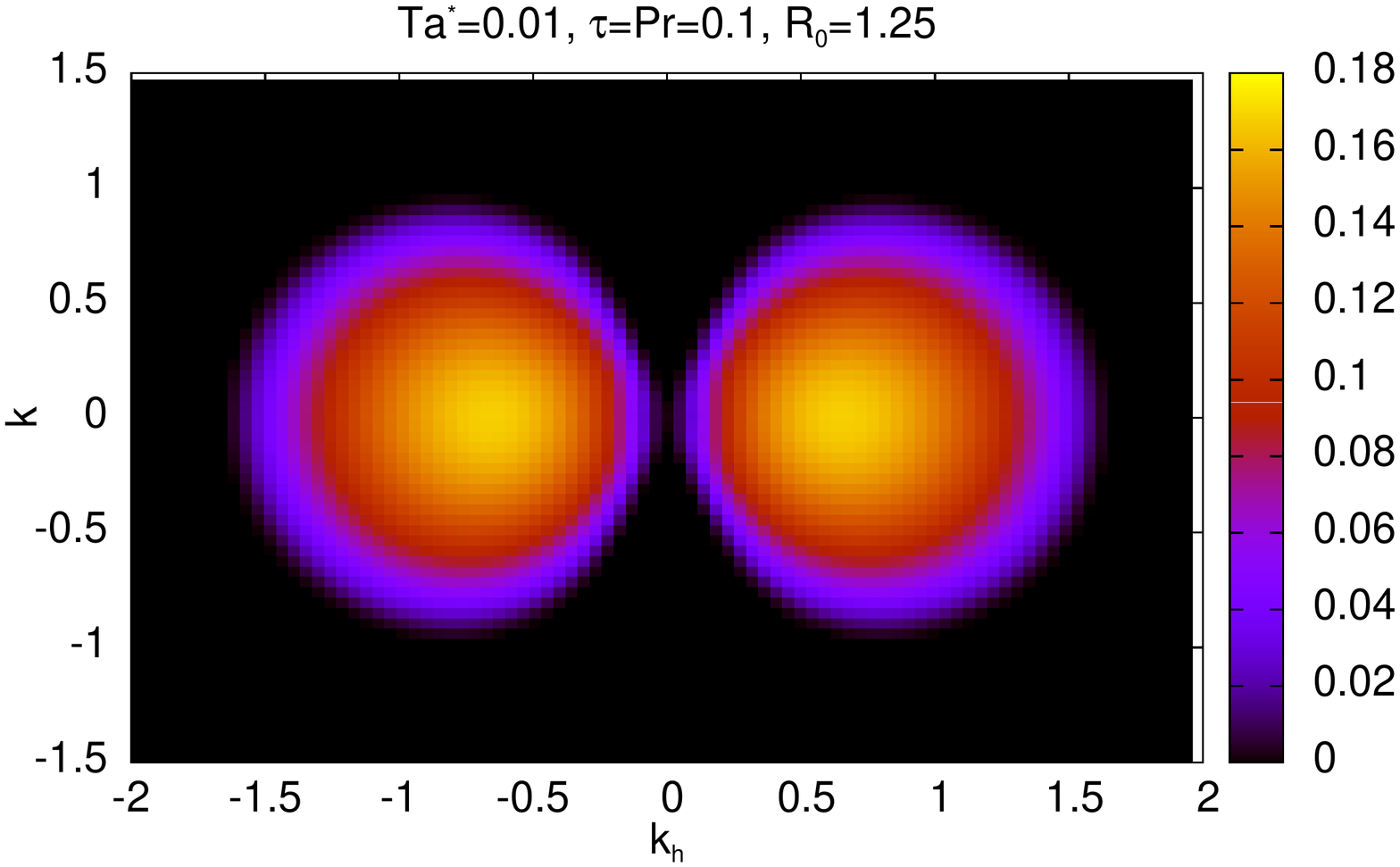}
\par\end{flushleft}%
\end{minipage}\hfill{}%
\begin{minipage}[t]{0.45\columnwidth}%
\begin{flushleft}
\includegraphics[scale=0.33]{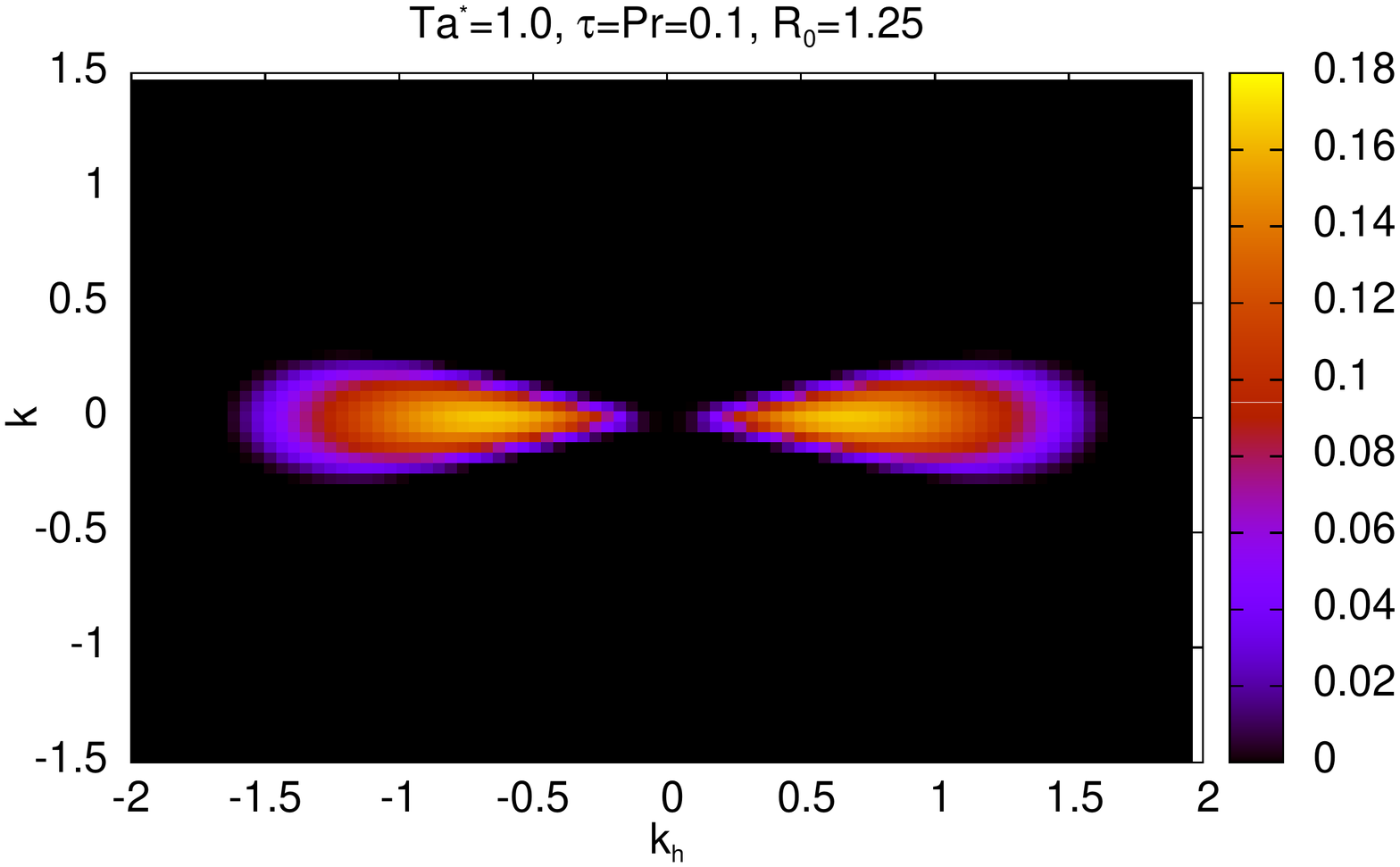}
\par\end{flushleft}%
\end{minipage}\caption{\label{fig:Fastest-growing-modes}Variation of the real part of $\lambda$
with input wavenumbers ($k,k_{h}$) for ${\rm Ta^{*}}=0.01$ (left)
and ${\rm Ta^{*}}=1$ (right) at $\tau={\rm Pr}=0.1$, $R_{0}=1.25$.
In both cases, $\theta=0$. Note that, as discussed in the text, the
fastest-growing modes have $k=0$.}
\end{figure}

Eq~\eqref{eq:dispersionrel} can be solved numerically for the growth
rate ($\lambda$) of the instability. The results for the rotating
case are shown in Fig~\ref{fig:Fastest-growing-modes} for ${\rm Ta^{*}}=0.01\text{ and }1$,
for $\tau={\rm Pr}=0.1$, $R_{0}=1.25$. It illustrates that, for
$\theta=0$, the fastest growing modes are those with $k=0$, as is
found for the non-rotating fingering unstable modes (\cite{2013ApJ...768...34B}).
These so called ``elevator'' modes are unaffected by rotation as
can be seen in Fig~\ref{fig:Fastest-growing-modes} and by direct
inspection of Eq~\eqref{eq:dispersionrel} since the last term (containing
${\rm Ta}^{*}$) vanishes for $k=0$ and $\theta=0$. The modes with
$k\neq0$ by contrast are suppressed by rotation in the sense that
the higher $k$ modes grow more slowly or become stable with increasing
${\rm Ta}^{*}$.\\
While linear theory helps to determine the linearly unstable regions
of parameter space, quantifying mixing by fingering convection can
only be done using nonlinear arguments. In the non-rotating case,
\cite{2012JFM...692....5R,2013ApJ...768...34B} showed that the nonlinear
saturation of the fingering instability is due to the shear that inevitably
develops between upflowing and downflowing fingers. By matching the
growth rates of the fingers to the growth rates of the emerging shear
instability, they successfully predicted the amplitude of the vertical
velocity at saturation which they then used to model the turbulent
mixing coefficient. 

Since rotation has a tendency to stabilize a system against motion
perpendicular to the rotation axis, we may expect it to stabilize
the fingers against the shear instabilities that cause their nonlinear
saturation. In that case, the vertical velocity within the fingers
might be permitted to grow to much larger amplitude before the secondary
shear instabilities develop, which could in turn lead to the enhancement
in the efficiency of vertical transport in rotating fingering convection
compared with the non-rotating case. This intuitive picture, and its
obvious potential for explaining the ``missing mixing'' in RGB stars,
motivated us to run DNSs of rotating fingering convection. In what
follows, we first attempt to estimate when the effects of rotation
may become important, and then present nonlinear DNSs of rotating
fingering convection to test these ideas. 

\section{Estimating when rotation is important in stellar interiors\label{sec:Estimating-when-rotation-is-important}}

While rotation does not have any effect on the growth rate of the
fastest-growing fingering modes, it is very likely to have one on
their nonlinear saturation (see our discussion above and the findings
of \cite{2017ApJ...834...44M} for the effect of rotation on the nonlinear
saturation of the ODDC instability). A commonly-used measure of the
relative strength of inertial forces ($\mathbf{u}\cdot\nabla\mathbf{u}$)
to Coriolis forces ($2\mathbf{\Omega}\times\mathbf{u}$) is the Rossby
number, defined as
\begin{equation}
{\rm Ro}=\frac{U}{2\Omega L},\label{eq:rossby-number}
\end{equation}
where $U$ and $L$ are typical dimensional velocities and length-scales
associated with the fluid motions in consideration. In turbulent flows,
the effect of rotation is therefore negligible if ${\rm Ro}\gg1$,
but dominant if ${\rm Ro}\ll1$. For moderate and high ${\rm Pr}$
fingering convection and ODDC, since $U\sim\frac{\kappa_{T}}{d}$
and $L\sim d$ (\cite{2011JFM...677..530T,2012ApJ...750...61M,2013ApJ...768..157W,2016ApJ...823...33M}),
one may estimate ${\rm Ro}$ as
\begin{equation}
{\rm Ro}\sim\frac{\frac{\kappa_{T}}{d}}{2\Omega d}\sim\frac{1}{\sqrt{{\rm Ta}^{*}}},\label{eq:rossbytaylorhighPr}
\end{equation}
which would imply that ${\rm Ta}^{*}\gg1$ double-diffusive systems
should be strongly rotationally constrained, while ${\rm Ta}^{*}\ll1$
systems should not feel the effect of rotation at all. This was verified
to be true for ${\rm Pr}\sim1$ down to ${\rm Pr}\sim0.01$ for ODDC
(see \cite{2017ApJ...834...44M}), for instance.\\
However in stellar interiors, the Prandtl number is asymptotically
small, taking values ranging from $10^{-6}$ down to $10^{\ensuremath{-9}}$.
In this regime, the vertical velocities within individual fingers
do not scale as above, but instead are expected to scale with Pr (\cite{2013ApJ...768...34B},
and see below). Hence, the effective Rossby number of rotating fingering
convection is predicted to be significantly different from the estimate
given in \eqref{eq:rossbytaylorhighPr}. \\
Indeed, for the parameter regime appropriate for stellar interiors,
$U$ and $L$ can be estimated using the results of \cite{2013ApJ...768...34B}.
They argue that
\begin{equation}
U\sim\frac{\lambda_{max}}{L},\ L\sim\frac{2\pi}{l_{max}},\label{eq:Brownetalmodel}
\end{equation}
where $\lambda_{max}$ is the growth rate of the fastest growing linearly
unstable mode, and $l_{max}$ is the associated horizontal wavenumber.
\begin{figure}[h]
\begin{centering}
\includegraphics[width=0.5\textwidth]{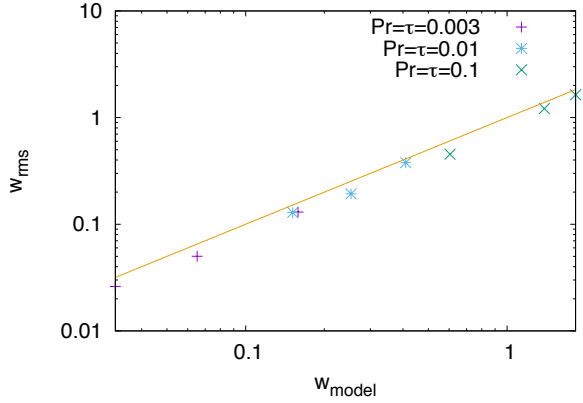}
\par\end{centering}
\caption{\label{fig:Comparison-of-results}Comparison of the rms vertical velocities
(symbols) estimated using the prediction of the Brown model (given
by the line representing $w_{model}=\frac{2\pi\lambda_{max}}{l_{max}}$)
with results of non-rotating DNSs from \cite{2011JFM...677..530T,2013ApJ...768...34B,2018AnRFM..50..275G}.}
\end{figure}
We can test this scaling by comparing for instance the r.m.s. vertical
velocity $w_{rms}$ extracted by reanalyzing non-rotating DNSs at
moderately low values of ${\rm Pr}$ presented in \cite{2011JFM...677..530T};
\cite{2013ApJ...768...34B,2018AnRFM..50..275G}, against our theoretical
prediction from \eqref{eq:Brownetalmodel}, namely,
\begin{equation}
w_{model}=\frac{\lambda_{max}l_{max}}{2\pi}
\end{equation}
where $\lambda_{max}$ and $l_{max}$ are found numerically by maximizing
the solutions of \eqref{eq:dispersionrel} for ${\rm Ta^{*}=0}$,
$k=0$ and $\theta=0$ against all possible value of $l$. This comparison
is shown in Fig \ref{fig:Comparison-of-results}, and clearly demonstrates
that $w_{model}$ is a remarkably accurate estimate for $w_{rms}$
across the entire range of Prandtl numbers and density ratios tested.
This would in turn imply that the Rossby number of rotating fingering
convection could, at a first approximation, be given by
\begin{equation}
{\rm Ro}=\frac{\lambda_{max}}{2\Omega}.
\end{equation}
In the asymptotic regime where ${\rm Pr},\tau\ll r\ll1$, where $r=\frac{R_{0}-1}{\tau^{-1}-1}$
(which is the regime most appropriate for stellar interiors), \cite{2013ApJ...768...34B}
further showed that 
\begin{equation}
\lambda_{max}\simeq{\rm Pr}\sqrt{\frac{\tau}{r{\rm Pr}}},
\end{equation}
resulting in the following predicted scaling for ${\rm Ro}$ with
all the input parameters:
\begin{equation}
{\rm Ro}\sim\sqrt{\frac{{\rm Pr}}{R_{0}-1}}\frac{1}{\sqrt{{\rm Ta}^{*}}}.\label{eq:Rossbyscaling}
\end{equation}
Since $d$ is related to the buoyancy frequency, $N$, as
\begin{equation}
d^{4}=\frac{\kappa_{T}\nu}{N^{2}},
\end{equation}
we can write ${\rm Ta}^{*}$ (given by \ref{eq:parameters}) as
\begin{equation}
{\rm Ta}^{*}={\rm Pr}\frac{\varOmega^{2}}{N^{2}}.
\end{equation}
Thus, our estimate for the Rossby number is simply given by
\begin{equation}
{\rm Ro}\sim\sqrt{\frac{N^{2}}{\varOmega^{2}}\frac{1}{R_{0}-1}}.
\end{equation}
Using a typical value of $N^{2}\sim10^{-4}$ in the radiative zone
of a solar-mass RGB star, one can estimate the Rossby number in the
region just above the hydrogen-burning shell using observed estimates
for red-giant rotation rates inferred from astroseismic data from
Kepler (\cite{2014A&A...564A..27D}) that range between $\sim0.25-10$
times the solar rotation rate ($\sim400nHz$). Assuming a rather extreme
estimate for density ratio $R_{0}\sim10^{3}$(\cite{2010ApJ...723..563D}\footnote{in reality, we should expect the density ratio to get larger in a
region where fingering convection acts to decrease the $\text{\textmu}-$gradient.}), we find that the Rossby number would be in the range $0.1\lesssim{\rm Ro}\lesssim10$
for slow rotators and $0.016\lesssim{\rm Ro}\lesssim1.6$ for fast
rotators, at the onset of fingering convection in RGB stars. This
then strongly suggests that rotation must be taken into account in
modeling fingering convection in these objects.\\
In the following section, we therefore present new DNSs of rotating
fingering convection, with values of ${\rm Ro}$ spanning the anticipated
range ($0.05-5$) for RGB stars.

\section{Numerical simulations\label{sec:Numerical-simulations}}

\subsection{Numerical tool: PADDI}

We use a version of the pseudo-spectral, triply periodic PADDI code
(\cite{2008GGG.....9.5003S}, \cite{2011JFM...677..530T}, \cite{2011JFM...677..554S})
modified in \cite{2017ApJ...834...44M} to include the effects of
rotation. We perform DNSs for ${\rm Ta}^{*}=0,0.01,0.1,1,10,25$ and
$100$. We anticipate fingers to become taller for increasing values
of $R_{0}\text{ or }{\rm Ta}^{*}$ and hence choose an elongated (rectangular)
box with dimensions $100d\times100d\times200d$ as our default domain
size. This is adjusted as and when required for varying $R_{0}$ and
${\rm Ta}^{*}$ (see Table 1). For simplicity, we only present results
for a domain at the poles with the rotation axis aligned with the
\textit{z}-direction, i.e. $\theta=0$ in \eqref{eq:omegaunitvector}.
In all of our simulations, the temperature and composition fields
are initialized with small amplitude random noise. Since performing
DNSs at realistic values of ${\rm Pr},\tau$ for stellar interiors
is computationally unfeasible as of now, we can only run simulations
at parameters down to ${\rm Pr}=\tau=0.01$ at best. For this exploratory
work, we prefer ${\rm Pr}=\tau=0.1$, because it allows us to comprehensively
explore the effects of varying $R_{0}$ and ${\rm Ta}^{*}$. We now
look at a few sample simulations. 

\subsection{Sample runs at $Pr=\tau=0.1$}

\begin{figure}
\begin{centering}
\includegraphics[width=1\textwidth]{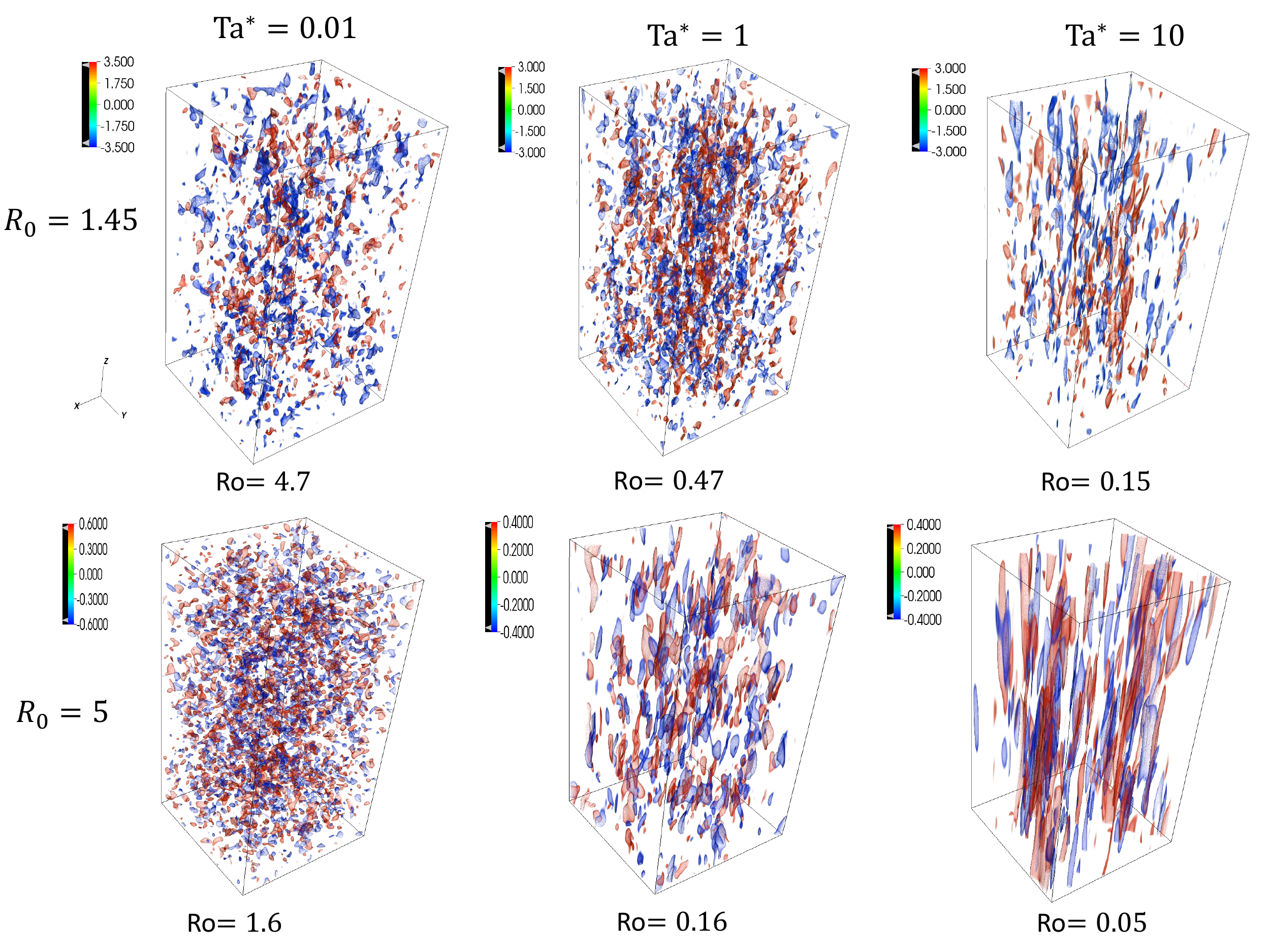}
\par\end{centering}
\caption{\label{fig:Snapshots-of-verticalvel}Snapshots of vertical velocity
fields (after saturation) at ${\rm Ta}^{*}=0.01$ (top), $1$ (middle)
and $10$ (bottom) for $R_{0}=1.45$ (left) and $5$ (right).}
\end{figure}

\begin{figure}
\begin{centering}
\includegraphics[width=0.5\textwidth]{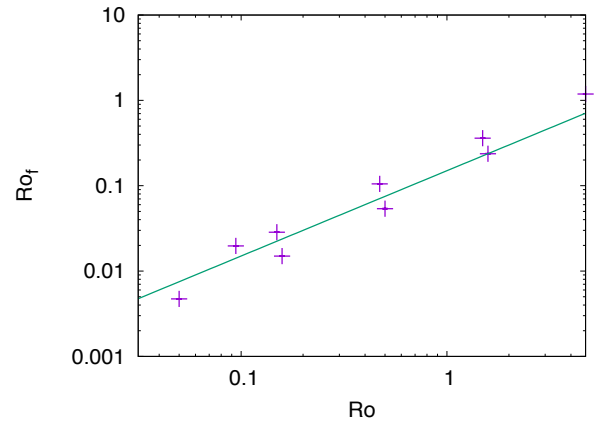}
\par\end{centering}
\caption{\label{fig:Comparison-of-Rossby}Comparison of the predicted Rossby
number, ${\rm Ro}$ (given by \ref{eq:Rossbyscaling}) with the effective
Rossby number of the fingers, ${\rm Ro}_{f}$, measured from the DNSs
according to \eqref{eq:Rossbyfingers}.}

\end{figure}

For this choice of $\tau=0.1$, a system is unstable to fingering
provided $1<R_{0}<10$. We focus our study on two values of $R_{0}=1.45$
and $5$ - the former representing conditions close to being convectively
unstable and the latter being half-way through the fingering-unstable
range. We summarize the results of our DNSs for different choices
of ${\rm Ta^{*}}$ (which varies with the Rossby number ${\rm Ro}$)
and $R_{0}$ in Table~1. \\
Fig~\ref{fig:Snapshots-of-verticalvel} shows snapshots of the vertical
velocity field in six different simulations spanning values of ${\rm Ta}^{*}=0.01,1\,\mbox{and\,}10$
for two values of $R_{0}=1.45$ and $5$. As can be readily seen from
the snapshots, at ${\rm Ta}^{*}=0.01$ (which is in the ``slowly
rotating'' regime), the fingers become more stable with increasing
stratification (i.e. increasing $R_{0}$) which has also been observed
experimentally (\cite{2003JFM...483..287K}) and in DNSs of non-rotating
fingering convection (\cite{2011JFM...677..530T}). With increasing
values of ${\rm Ta}^{*}$, we observe a propensity of the flow to
become invariant along the axis of rotation. This is in accordance
with the Taylor-Proudman theorem (\cite{1916RSPSA..92..408P,1917RSPSA..93...99T}),
which becomes relevant when the Rossby number becomes much smaller
than $1$. The Taylor-Proudman constraint is significantly more pronounced
for the $R_{0}=5$ case than for the $R_{0}=1.45$ case, at fixed
${\rm Ta}^{*}$. To understand why this is the case, we note that
the effective Rossby number, given by \eqref{eq:Rossbyscaling}, is
significantly higher at $R_{0}=1.45$ (${\rm Ro}=0.47$ for ${\rm Ta}^{*}=1$)
than $R_{0}=5$ (${\rm Ro}=0.16$ for ${\rm Ta}^{*}=1$); hence achieving
a Taylor-Proudman state at smaller $R_{0}$ requires larger values
of ${\rm Ta^{*}}$.\\
\begin{table}
\begin{centering}
\begin{tabular}{|c|c|c|c|c|c|c|}
\hline 
{\footnotesize{}$R_{0}$} & {\footnotesize{}${\rm Ta}^{*}$} & {\footnotesize{}Resolution}\footnote{{\footnotesize{}Note that the resolution is given in terms of number
of Fourier modes used. }} & {\footnotesize{}$L_{x}\times L_{y}\times L_{z}$} & {\footnotesize{}${\rm Nu}{}_{\mu}$} & {\footnotesize{}$w_{rms}$} & {\footnotesize{}$u_{rms}$}\tabularnewline
\hline 
\hline 
{\footnotesize{}$1.45$} & {\footnotesize{}$0.0$} & {\footnotesize{}$96\times96\times96$} & {\footnotesize{}$100\times100\times100$} & {\footnotesize{}$86.415\pm2.378$} & {\footnotesize{}$1.2095\pm0.0135$} & {\footnotesize{}$1.5615\pm0.0169$}\tabularnewline
 & \textbf{\footnotesize{}$0.01$} & \textbf{\footnotesize{}$64\times64\times128$} & \textbf{\footnotesize{}$100\times100\times200$} & \textbf{\footnotesize{}$83.395\pm1.196$} & \textbf{\footnotesize{}$1.187\pm0.0006$} & \textbf{\footnotesize{}$1.544\pm0.006$}\tabularnewline
 & {\footnotesize{}$0.1$} & {\footnotesize{}$64\times64\times128$} & {\footnotesize{}$100\times100\times200$} & {\footnotesize{}$75.497\pm0.468$} & \textbf{\footnotesize{}$1.1406\pm0.0039$} & {\footnotesize{}$1.5304\pm0.0045$}\tabularnewline
 & {\footnotesize{}$1$} & {\footnotesize{}$64\times64\times128$} & {\footnotesize{}$100\times100\times200$} & {\footnotesize{}$62.403\pm0.857$} & \textbf{\footnotesize{}$1.058\pm0.006$} & {\footnotesize{}$1.482\pm0.008$}\tabularnewline
 &  & {\footnotesize{}$128\times128\times256$} & {\footnotesize{}$100\times100\times200$} & {\footnotesize{}$61.644\pm1.843$} & {\footnotesize{}$1.0507\pm0.0158$} & {\footnotesize{}$1.476\pm0.021$}\tabularnewline
 & \textcolor{black}{\footnotesize{}$10$} & \textcolor{black}{\footnotesize{}$64\times64\times128$}{\footnotesize{}}\footnote{{\footnotesize{}This run emerged with a cyclonic large scale vortex
(see Section \ref{subsec:Emergence-of-LSV}) }} & \textcolor{black}{\footnotesize{}$100\times100\times200$} & \textcolor{black}{\footnotesize{}$1899.28\pm158.055$} & \textcolor{black}{\footnotesize{}$2.209\pm0.005$} & \textcolor{black}{\footnotesize{}$7.31\pm0.16$}\tabularnewline
 &  & {\footnotesize{}$128\times128\times256$} & {\footnotesize{}$100\times100\times200$} & {\footnotesize{}$44.186\pm1.49$} & {\footnotesize{}$0.9012\pm0.0128$} & {\footnotesize{}$1.495\pm0.048$}\tabularnewline
 & {\footnotesize{}$25$} & {\footnotesize{}$64\times64\times128$} & {\footnotesize{}$100\times100\times200$} & {\footnotesize{}$53.26\pm3.294$} & {\footnotesize{}$0.984\pm0.034$} & {\footnotesize{}$1.5859\pm0.0298$}\tabularnewline
 & {\footnotesize{}$100$} & {\footnotesize{}$64\times64\times128$} & {\footnotesize{}$100\times100\times200$} & {\footnotesize{}$57.65\pm4.07$} & {\footnotesize{}$1.028\pm0.047$} & {\footnotesize{}$1.7358\pm0.0431$}\tabularnewline
\hline 
{\footnotesize{}$5.0$} & {\footnotesize{}$0.0$} & {\footnotesize{}$128\times128\times128$}\footnote{data from \cite{2011JFM...677..530T} } & {\footnotesize{}$100\times100\times100$} & {\footnotesize{}$11.396\pm0.109$} & {\footnotesize{}$0.2575\pm0.0019$} & {\footnotesize{}$0.306\pm0.002$}\tabularnewline
 & \textbf{\footnotesize{}$0.01$} & {\footnotesize{}$64\times64\times128$} & {\footnotesize{}$100\times100\times200$} & {\footnotesize{}$9.838\pm0.1545$} & {\footnotesize{}$0.23704\pm0.0022$} & {\footnotesize{}$0.2929\pm0.0022$}\tabularnewline
 & {\footnotesize{}$1$} & {\footnotesize{}$32\times32\times64$} & {\footnotesize{}$100\times100\times200$} & {\footnotesize{}$4.649\pm0.137$} & {\footnotesize{}$0.1492\pm0.0029$} & {\footnotesize{}$0.2355\pm0.0039$}\tabularnewline
 &  & {\footnotesize{}$64\times64\times128$} & {\footnotesize{}$100\times100\times400$} & {\footnotesize{}$4.681\pm0.152$} & {\footnotesize{}$0.1496\pm0.0033$} & {\footnotesize{}$0.2397\pm0.0041$}\tabularnewline
 & {\footnotesize{}$10$} & {\footnotesize{}$64\times64\times128$} & {\footnotesize{}$100\times100\times200$} & {\footnotesize{}$4.986\pm0.260$} & {\footnotesize{}$0.1535\pm0.0049$} & {\footnotesize{}$0.2277\pm0.0036$}\tabularnewline
 &  & {\footnotesize{}$64\times64\times128$} & {\footnotesize{}$100\times100\times400$} & {\footnotesize{}$4.8046\pm0.3054$} & {\footnotesize{}$0.1494\pm0.0056$} & {\footnotesize{}$0.256\pm0.008$}\tabularnewline
 & {\footnotesize{}$100$} & {\footnotesize{}$64\times64\times128$} & {\footnotesize{}$100\times100\times200$} & {\footnotesize{}$6.112\pm0.301$} & {\footnotesize{}$0.167\pm0.005$} & {\footnotesize{}$0.193\pm0.006$}\tabularnewline
\hline 
{\footnotesize{}$9.1$} & {\footnotesize{}$0.0$} & {\footnotesize{}$32\times32\times64$} & {\footnotesize{}$100\times100\times200$} & {\footnotesize{}$1.3812\pm0.0142$} & {\footnotesize{}$0.02989\pm0.0006$} & {\footnotesize{}$0.0338\pm0.0006$}\tabularnewline
 & \textbf{\footnotesize{}$0.01$} & {\footnotesize{}$64\times64\times128$} & {\footnotesize{}$100\times100\times200$} & {\footnotesize{}$1.0254\pm0.0013$} & {\footnotesize{}$0.0073\pm0.0002$} & {\footnotesize{}$0.0122\pm0.0004$}\tabularnewline
 & {\footnotesize{}$0.1$} & {\footnotesize{}$32\times32\times64$} & {\footnotesize{}$100\times100\times200$} & {\footnotesize{}$1.016\pm0.001$} & {\footnotesize{}$0.0057\pm0.0002$} & {\footnotesize{}$0.01025\pm0.00029$}\tabularnewline
 & {\footnotesize{}$1$} & {\footnotesize{}$32\times32\times64$} & {\footnotesize{}$100\times100\times200$} & {\footnotesize{}$1.0267\pm0.0045$} & {\footnotesize{}$0.0069\pm0.0006$} & {\footnotesize{}$0.0107\pm0.0011$}\tabularnewline
 &  & {\footnotesize{}$32\times32\times64$} & {\footnotesize{}$100\times100\times800$} & {\footnotesize{}$1.0153\pm0.0011$} & {\footnotesize{}$0.0056\pm0.0002$} & {\footnotesize{}$0.01003\pm0.0003$}\tabularnewline
 & {\footnotesize{}$10$} & {\footnotesize{}$32\times32\times64$} & {\footnotesize{}$100\times100\times200$} & {\footnotesize{}$1.248\pm0.018$} & {\footnotesize{}$0.0207\pm0.0008$} & {\footnotesize{}$0.02366\pm0.00095$}\tabularnewline
 &  & {\footnotesize{}$32\times32\times128$} & {\footnotesize{}$100\times100\times1600$} & {\footnotesize{}$1.01376\pm0.00089$} & {\footnotesize{}$0.0055\pm0.0002$} & {\footnotesize{}$0.00964\pm0.0003$}\tabularnewline
 & {\footnotesize{}$25$} & {\footnotesize{}$32\times32\times64$} & {\footnotesize{}$100\times100\times200$} & {\footnotesize{}$1.601\pm0.027$} & {\footnotesize{}$0.032\pm0.001$} & {\footnotesize{}$0.034\pm0.001$}\tabularnewline
 &  & {\footnotesize{}$32\times32\times64$} & {\footnotesize{}$100\times100\times800$} & {\footnotesize{}$1.039\pm0.005$} & {\footnotesize{}$0.00831\pm0.0006$} & {\footnotesize{}$0.01229\pm0.00099$}\tabularnewline
\hline 
\end{tabular}
\par\end{centering}
\caption{DNS runs for chosen set of parameters ${\rm Ta}^{*}\mbox{ and }R_{0}$
at ${\rm Pr}=\tau=0.1$ }
\end{table}
Using the set of rotating DNSs, we can actually compare our theoretical
estimate for the Rossby number ${\rm Ro}$ (see \eqref{eq:Rossbyscaling}),
to the effective Rossby number of the simulations which is given by
\begin{equation}
{\rm Ro}_{f}\sim\frac{w_{rms}}{10\sqrt{{\rm Ta}^{*}}},\label{eq:Rossbyfingers}
\end{equation}
where $w_{rms}$ is the measured rms vertical velocity in the DNS,
and the number $10$ comes from assuming that the horizontal dimensions
of the fingers are of the order of $10d$ (which roughly corresponds
to the width of the fastest-growing fingers). The results are shown
in Fig \ref{fig:Comparison-of-Rossby} and confirm that the predicted
${\rm Ro}$ derived in Section 4 is a fairly good estimate of the
effective Rossby number of the fingers (${\rm Ro}_{f}$) in all of
our rotating simulations. 

Finally, note that as in \cite{2011PhDT........23T}, the elongation
of the fingers along the vertical direction (either for high $R_{0}$,
or high ${\rm Ta^{*}}$, or both) poses a numerical challenge. Indeed,
we need to ensure that our domain is large enough so that the fingers
do not ``feel'' its boundaries, which would lead to artificial enhancements
of the transport rates due to the assumption of periodic boundary
conditions. This problem is discussed in more detail in Appendix.

\subsection{Effect of rotation on compositional transport by small-scale fingering
convection}

\begin{figure}
\begin{raggedright}
\begin{minipage}[t]{0.45\columnwidth}%
\begin{flushleft}
\includegraphics[scale=0.33]{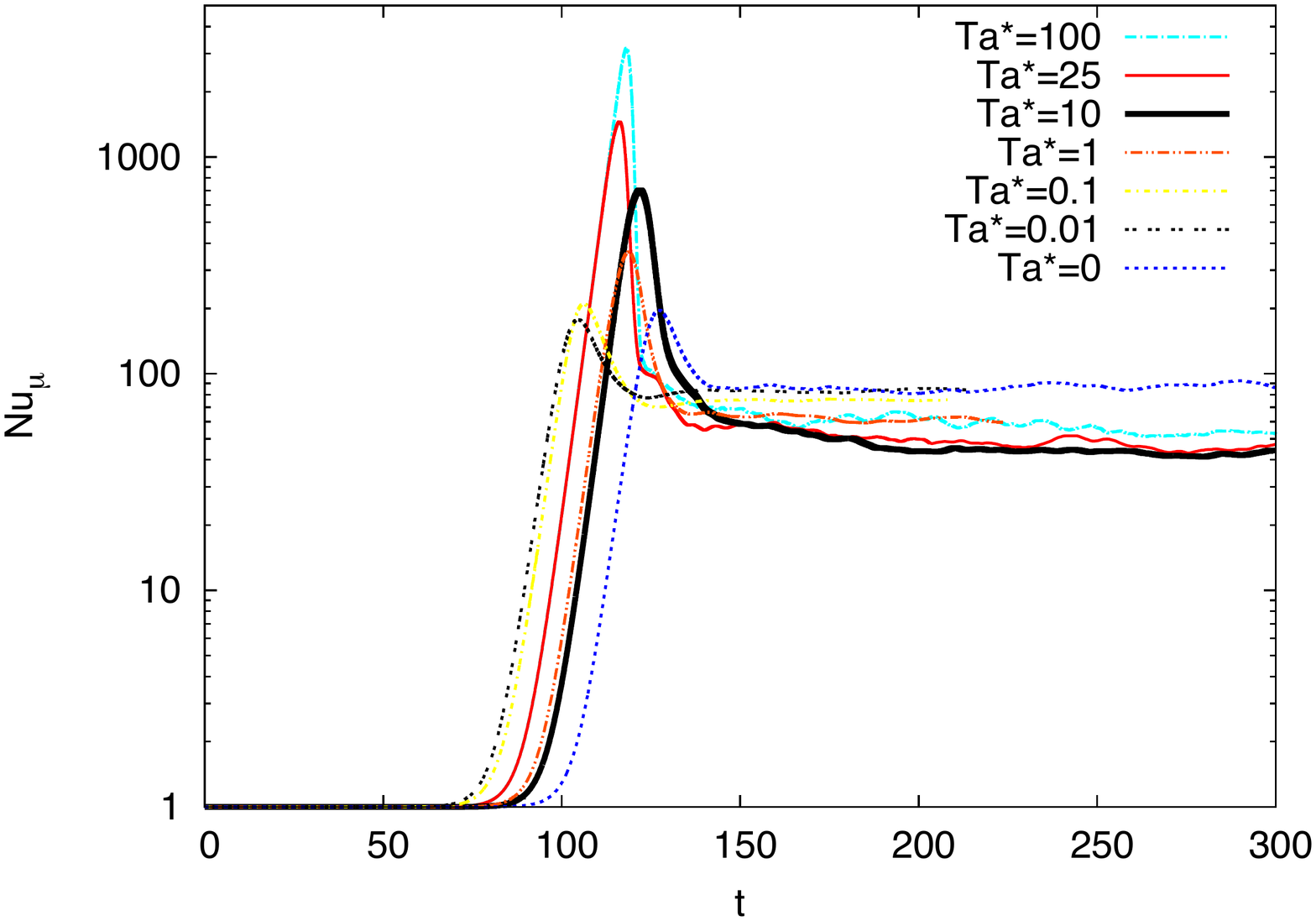}
\par\end{flushleft}%
\end{minipage}\hfill{}%
\begin{minipage}[t]{0.45\columnwidth}%
\begin{flushleft}
\includegraphics[scale=0.33]{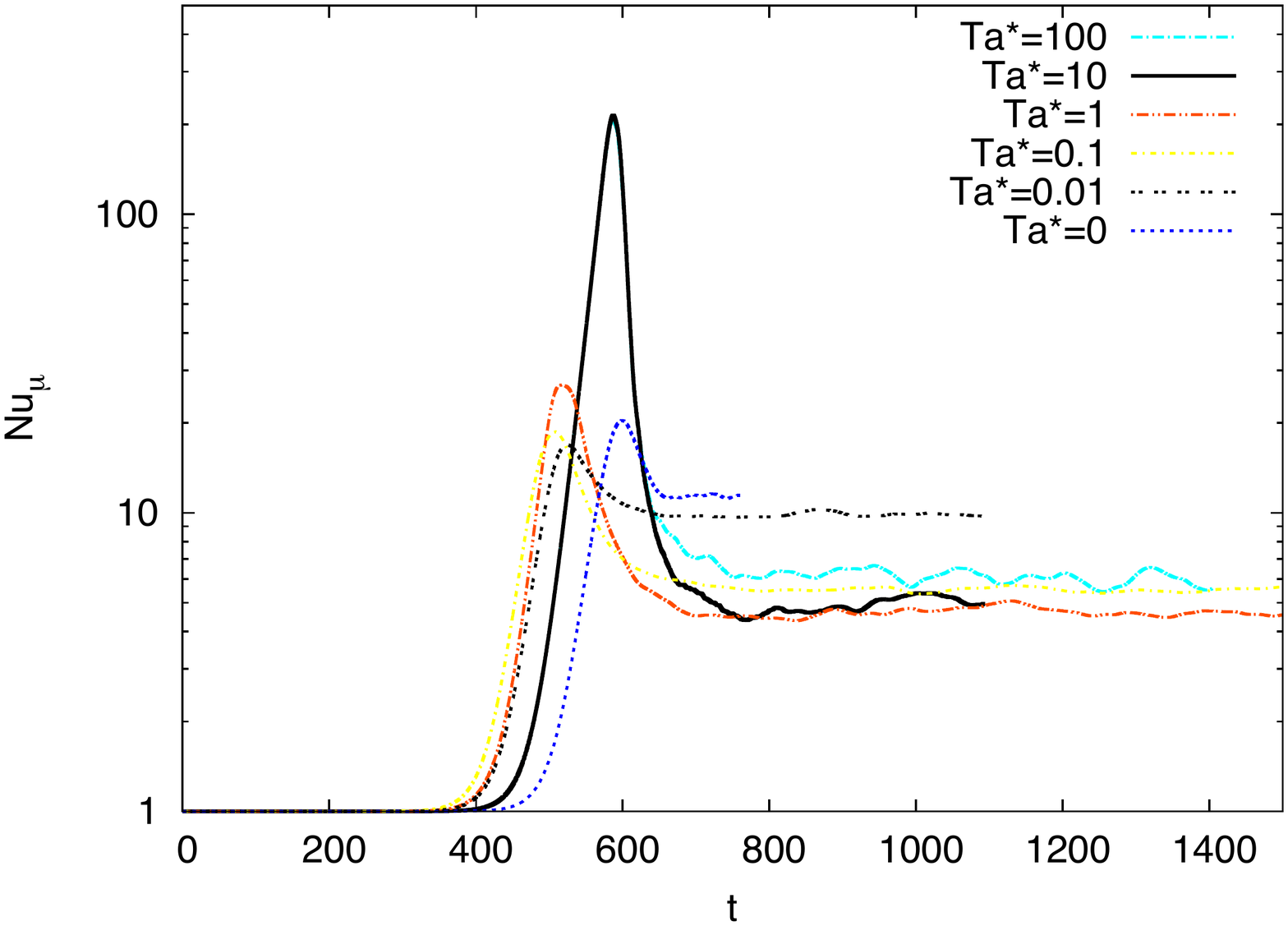}
\par\end{flushleft}%
\end{minipage}
\par\end{raggedright}
\caption{\label{fig:fluxes at saturation} ${\rm Nu}{}_{\mu}$ as function
of time (in units of diffusion time scale) for simulations with ${\rm Pr}=\tau=0.1$
, and varying ${\rm Ta}^{*}$ at $R_{0}=1.45$ (left) and $5$ (right).}
\end{figure}

\begin{figure}[H]
\begin{centering}
\includegraphics[width=0.5\textwidth]{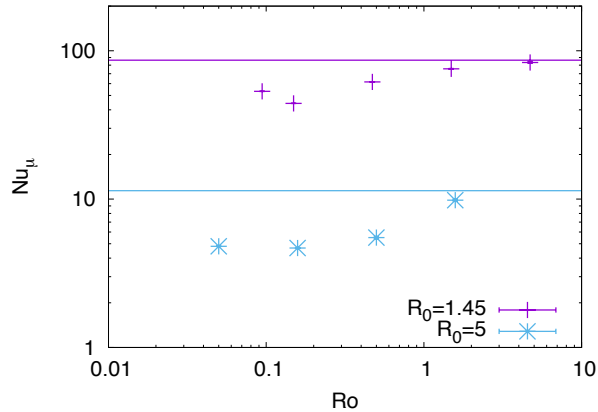}
\par\end{centering}
\caption{\label{fig:Time-averaged-fluxes}Time-averaged ${\rm Nu}{}_{\mu}$
in the statistically steady state as a function of the estimated Rossby
number ${\rm Ro}$ at $R_{0}=1.45$ and $5$ - the horizontal lines
shows the corresponding non-rotating (${\rm Ro}\rightarrow\infty$)
${\rm Nu}{}_{\mu}$ values.}
\end{figure}

\begin{figure}
\begin{raggedright}
\begin{minipage}[t]{0.45\columnwidth}%
\begin{flushleft}
\includegraphics[scale=0.3]{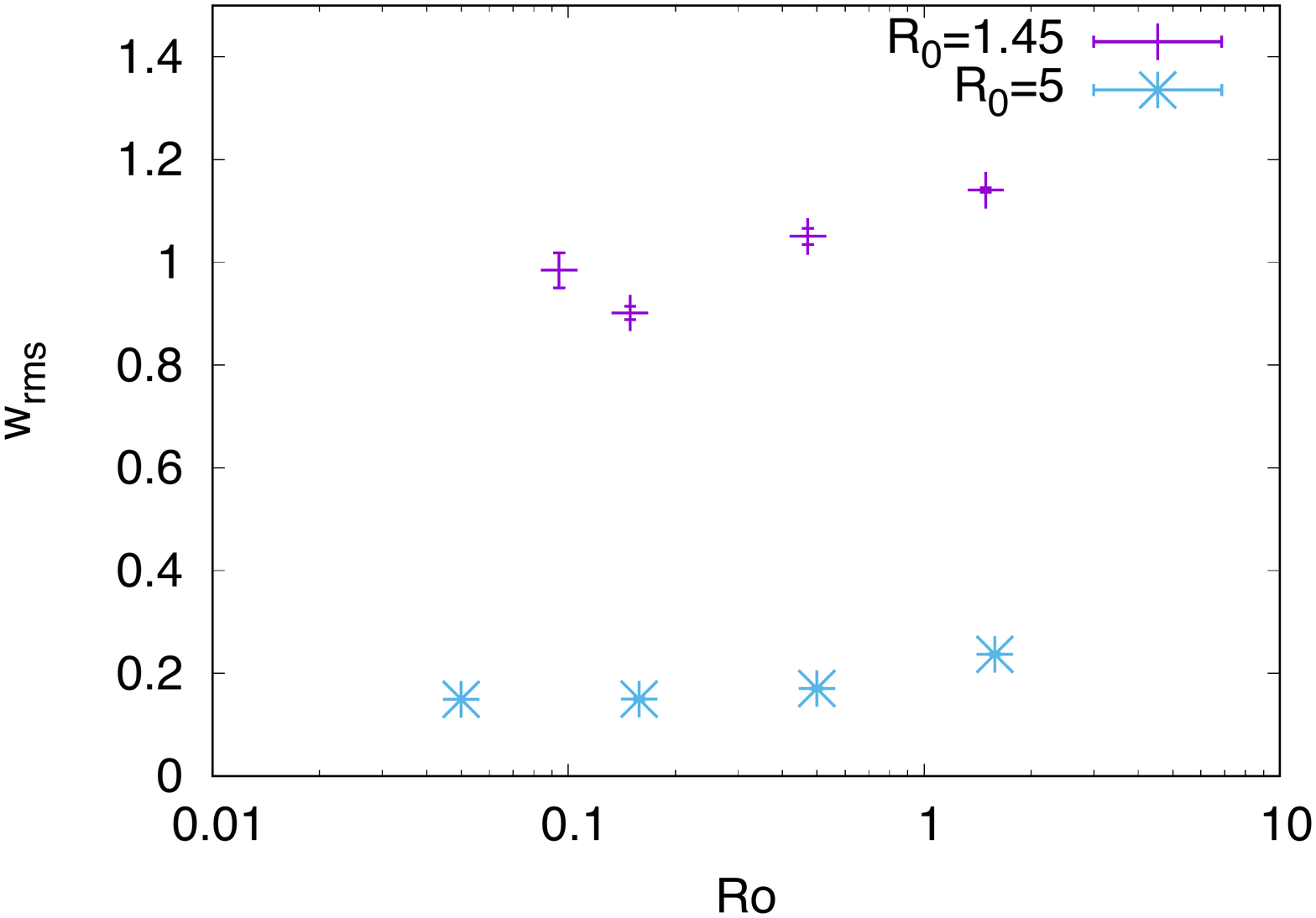}
\par\end{flushleft}%
\end{minipage}\hfill{}%
\begin{minipage}[t]{0.45\columnwidth}%
\begin{flushleft}
\includegraphics[scale=0.3]{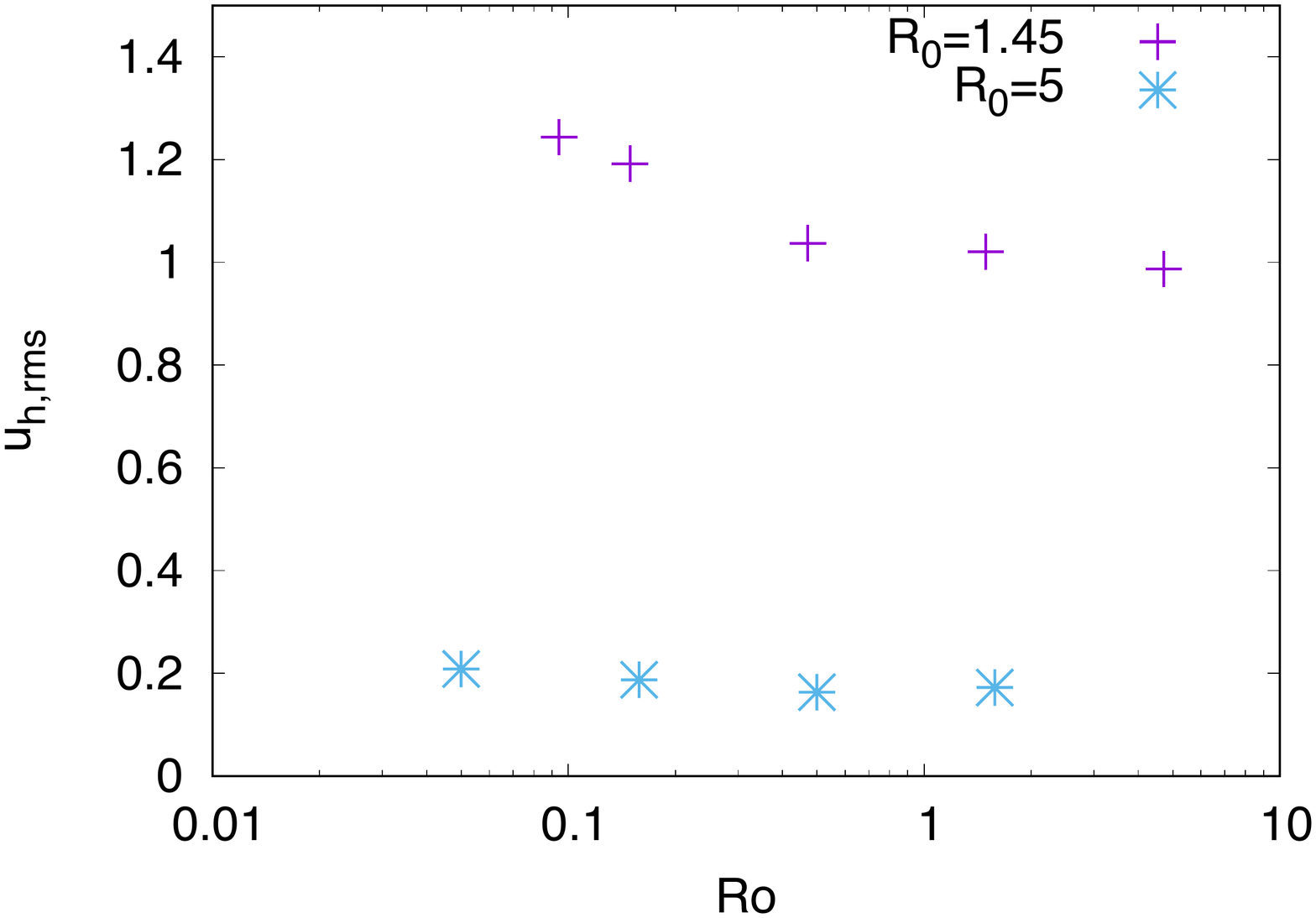}
\par\end{flushleft}%
\end{minipage}
\par\end{raggedright}
\caption{\label{fig:Ratio-of-the-velocities}The rms vertical ($w_{rms}$)
and total horizontal ($u_{h,rms}=\sqrt{u_{rms}^{2}+v_{rms}^{2}}$)
velocities as a function of the Rossby number ${\rm Ro}$ for DNSs
at $R_{0}=1.45$ and 5.}

\end{figure}

In what follows, we now only report on the simulations with the largest
resolution and domain sizes available at $R_{0}=1.45$ and $5$. We
measure the vertical flux of composition in terms of the compositional
Nusselt number, ${\rm Nu_{\mu}}$ defined as:
\begin{equation}
{\rm Nu_{\mu}}=1-\frac{R_{0}}{\tau}\langle\tilde{w}\tilde{\mu}\rangle,
\end{equation}
where $\left\langle \right\rangle $ denotes a volume average over
the entire domain. ${\rm Nu}_{\mu}$ can be interpreted as the ratio
of the effective diffusivity $D_{\mu}$ to the microscopic diffusivity
$\kappa_{\mu}$ i.e. 
\begin{equation}
D_{\mu}=\kappa_{\mu}{\rm Nu_{\mu}.}
\end{equation}
As usual, the turbulent transport of heat is negligible in fingering
convection. The time-evolution of ${\rm Nu_{\mu}}$ is shown in Figure~\ref{fig:fluxes at saturation}
for different values of ${\rm Ta}^{*}$. As expected, we see the development
of the fingering instability at early times, followed by its nonlinear
saturation. As anticipated from our naive argument of Section 3, we
find that the peak compositional transport increases significantly
with ${\rm Ta}{}^{*}$, suggesting that the shear instability between
the fingers is indeed stabilized by rotation. However, we also see
that the turbulent transport rates after saturation of the fingering
instability, once the system has achieved a statistically steady state,
do not depend on ${\rm Ta}{}^{*}$ nearly as much. To see this more
quantitatively, we measure the transport properties of fingering convection
in that statistically stationary state. 
The time-averaged ${\rm Nu}{}_{\mu}$ values thus extracted for different
values of ${\rm Ta}^{*}$ are presented in Figure~\ref{fig:Time-averaged-fluxes}
for both values of $R_{0}$, as a function of the corresponding Rossby
numbers, ${\rm Ro}$ (as given by Eq \ref{eq:Rossbyscaling}). This
shows that rotation actually tends to lower the vertical transport
rates by a factor of up to 2 compared with the non-rotating case.
This is a rather unexpected finding in light of our discussions in
Section 3 where we expected that rotation would act to enhance the
r.m.s. vertical velocities and therefore also the mixing rates. Instead,
we find that both vertical and horizontal r.m.s. velocities remains
almost unchanged as the rotation rate is increased (see Fig \ref{fig:Ratio-of-the-velocities}).

\subsection{Emergence of a large scale vortex\label{subsec:Emergence-of-LSV}}

\begin{figure}
\begin{raggedright}
\begin{minipage}[t]{0.45\columnwidth}%
\begin{flushleft}
\includegraphics[scale=0.32]{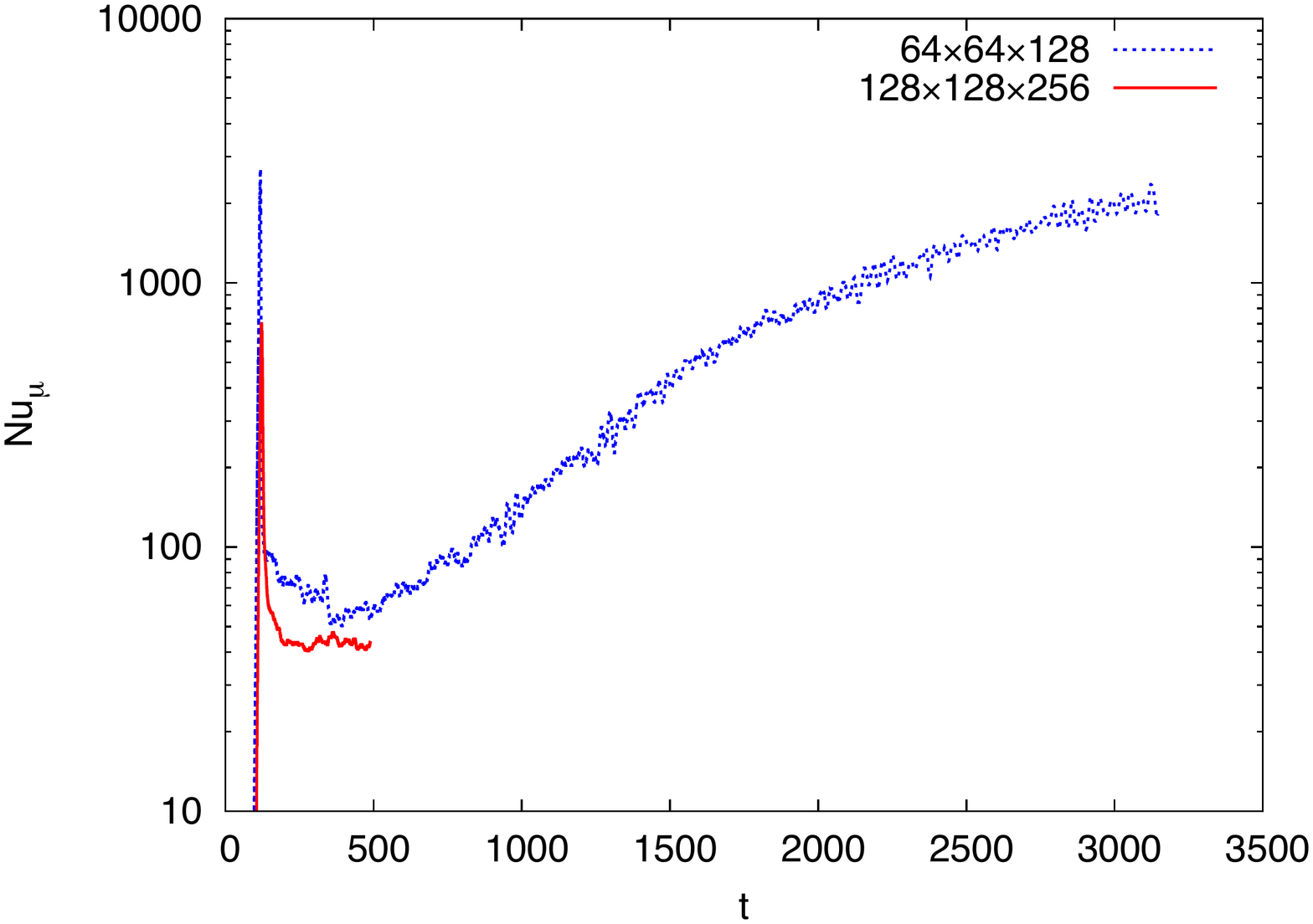}
\par\end{flushleft}%
\end{minipage}\hfill{}%
\begin{minipage}[t]{0.45\columnwidth}%
\begin{flushleft}
\includegraphics[scale=0.31]{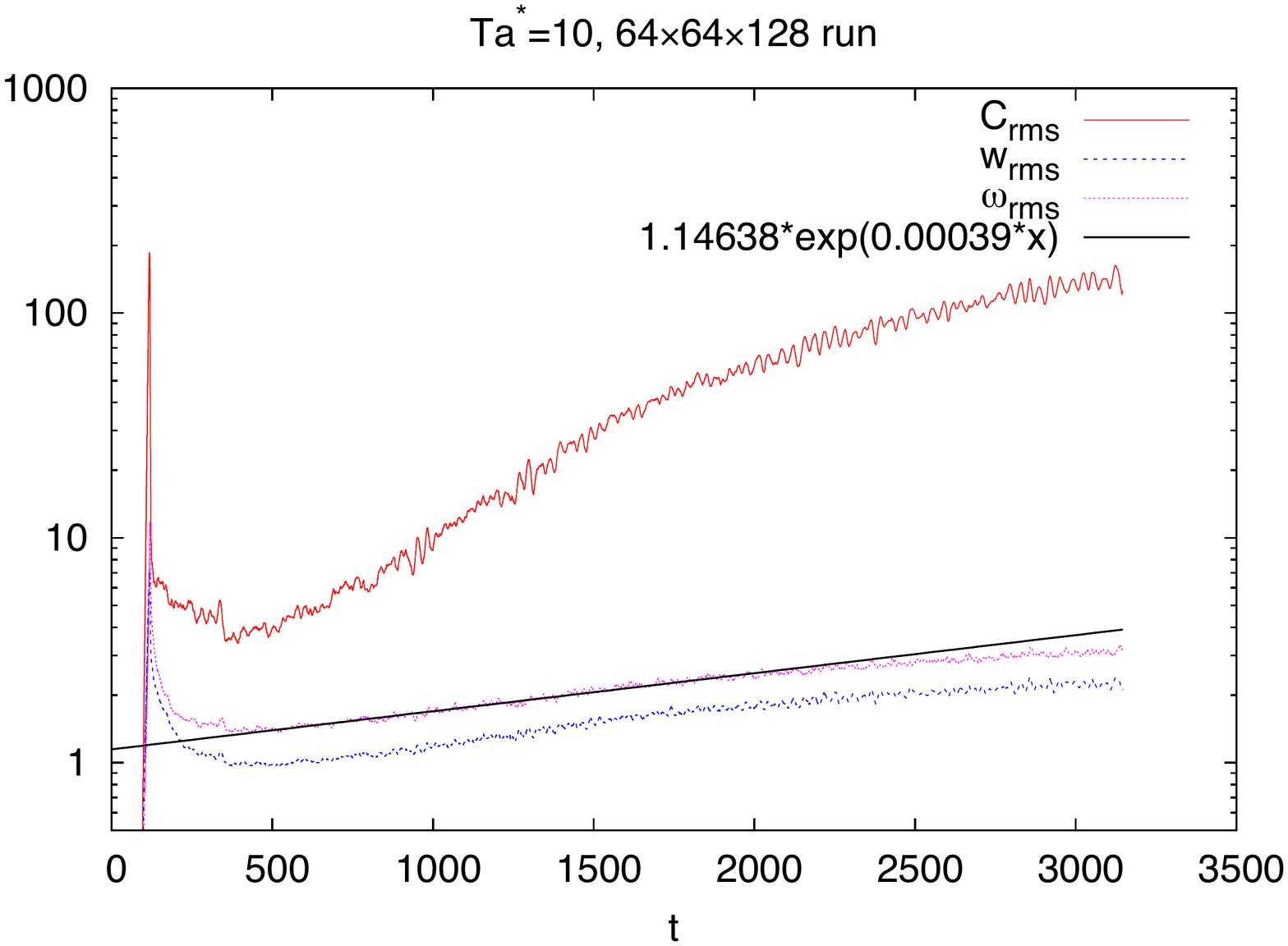}
\par\end{flushleft}%
\end{minipage}
\par\end{raggedright}
\caption{\label{fig:resolutioneffect}(left) Enhancement in chemical transport
rates, measured in terms of ${\rm Nu}_{\mu}$ observed at $R_{0}=1.45$
and ${\rm Ta}^{*}=10$ for {\small{}$64\times64\times128$ }run compared
to a {\small{}$128\times128\times256$} run; (right) growth in rms
values of the chemical field and vertical components of velocity and
vorticity in the {\small{}$64\times64\times128$ }run.}
\end{figure}

\begin{figure}
\begin{raggedright}
\includegraphics[scale=0.55]{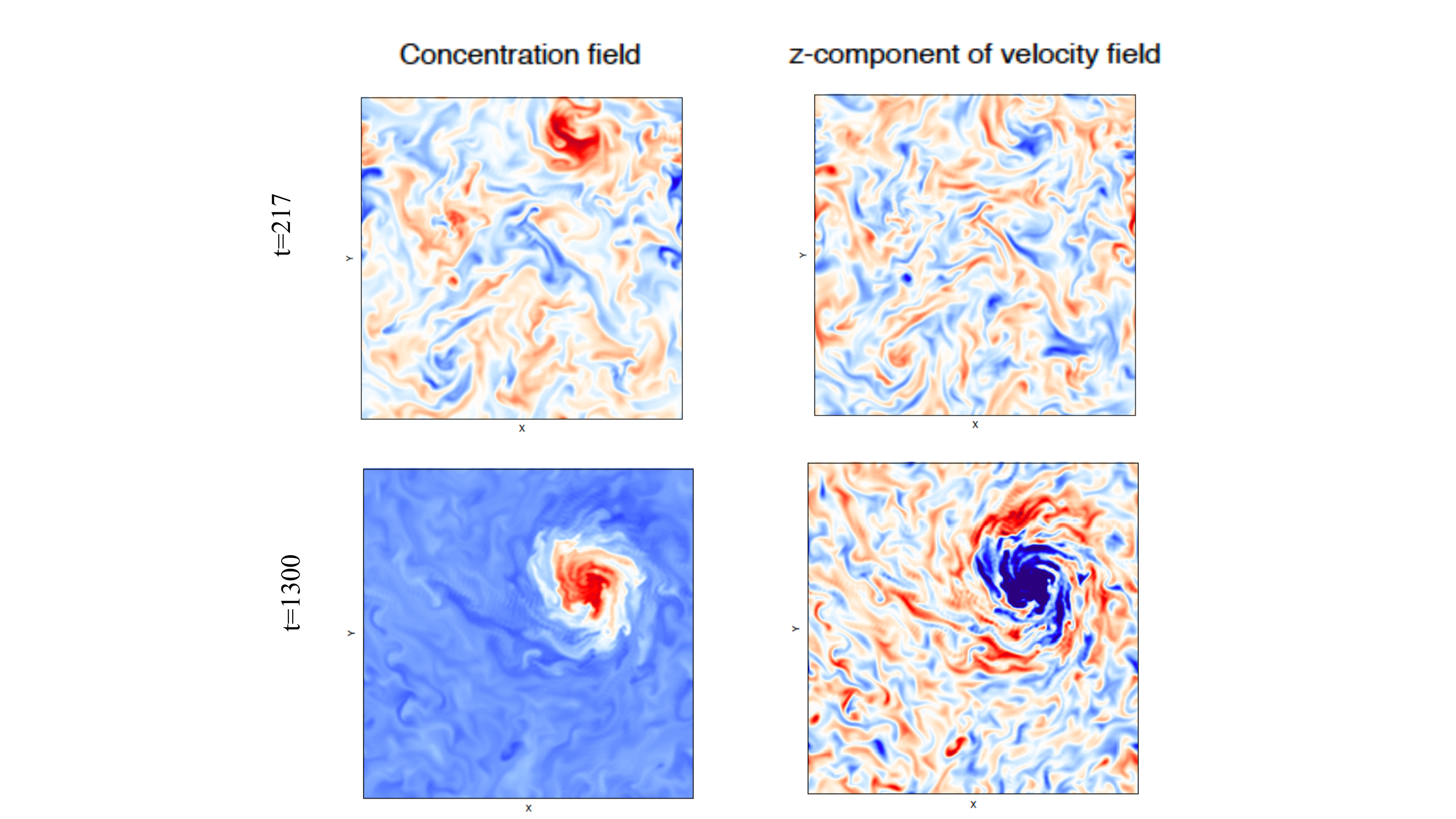}
\par\end{raggedright}
\caption{\label{fig:Snapshots-of-chemical_uz}Horizontal snapshots of chemical
(left) and vertical velocity (right) fields at $t=217$ (top) and
$t=1300$ (bottom) for the $64\times64\times128$ run at $R_{0}=1.45,\,{\rm Ta}^{*}=10$
- red shows positive and blue shows negative values of the quantities.}
\end{figure}

\begin{figure}
\begin{raggedright}
\begin{minipage}[t]{0.45\columnwidth}%
\begin{center}
\includegraphics[scale=0.4]{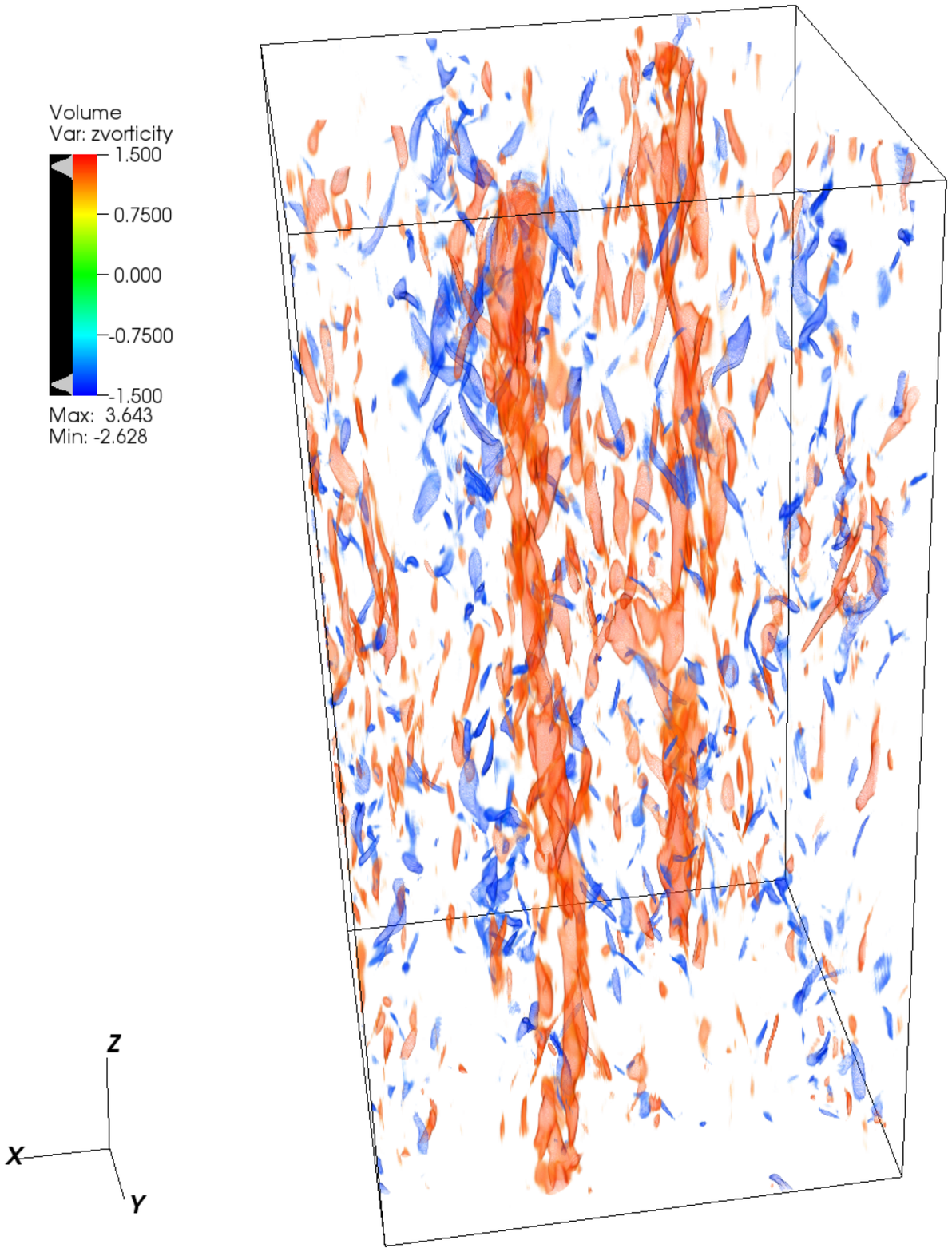}
\par\end{center}%
\end{minipage}\hfill{}%
\begin{minipage}[t]{0.45\columnwidth}%
\begin{center}
\includegraphics[scale=0.4]{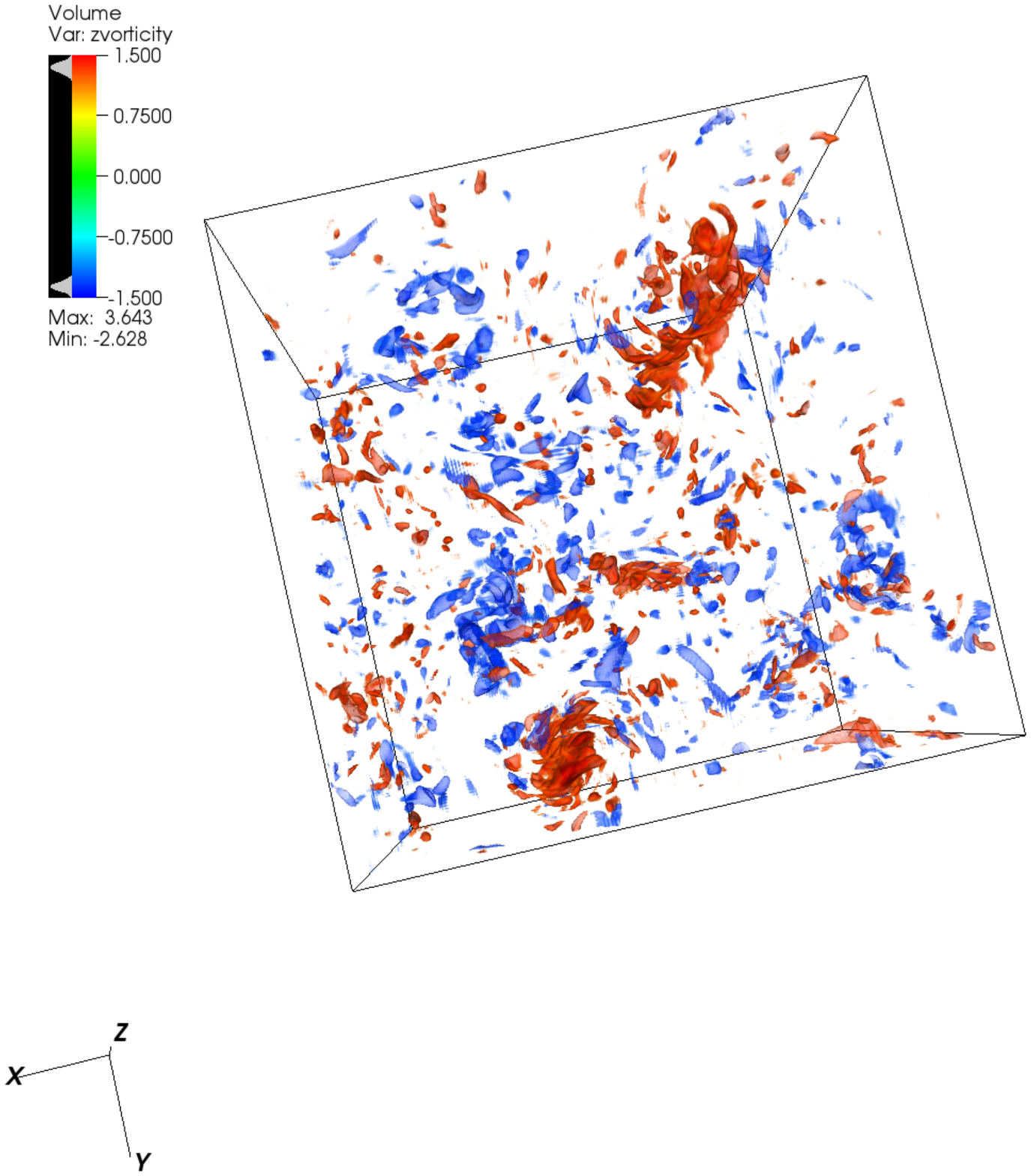}
\par\end{center}%
\end{minipage}
\par\end{raggedright}
\begin{raggedright}
\begin{minipage}[t]{0.45\columnwidth}%
\begin{center}
\includegraphics[scale=0.4]{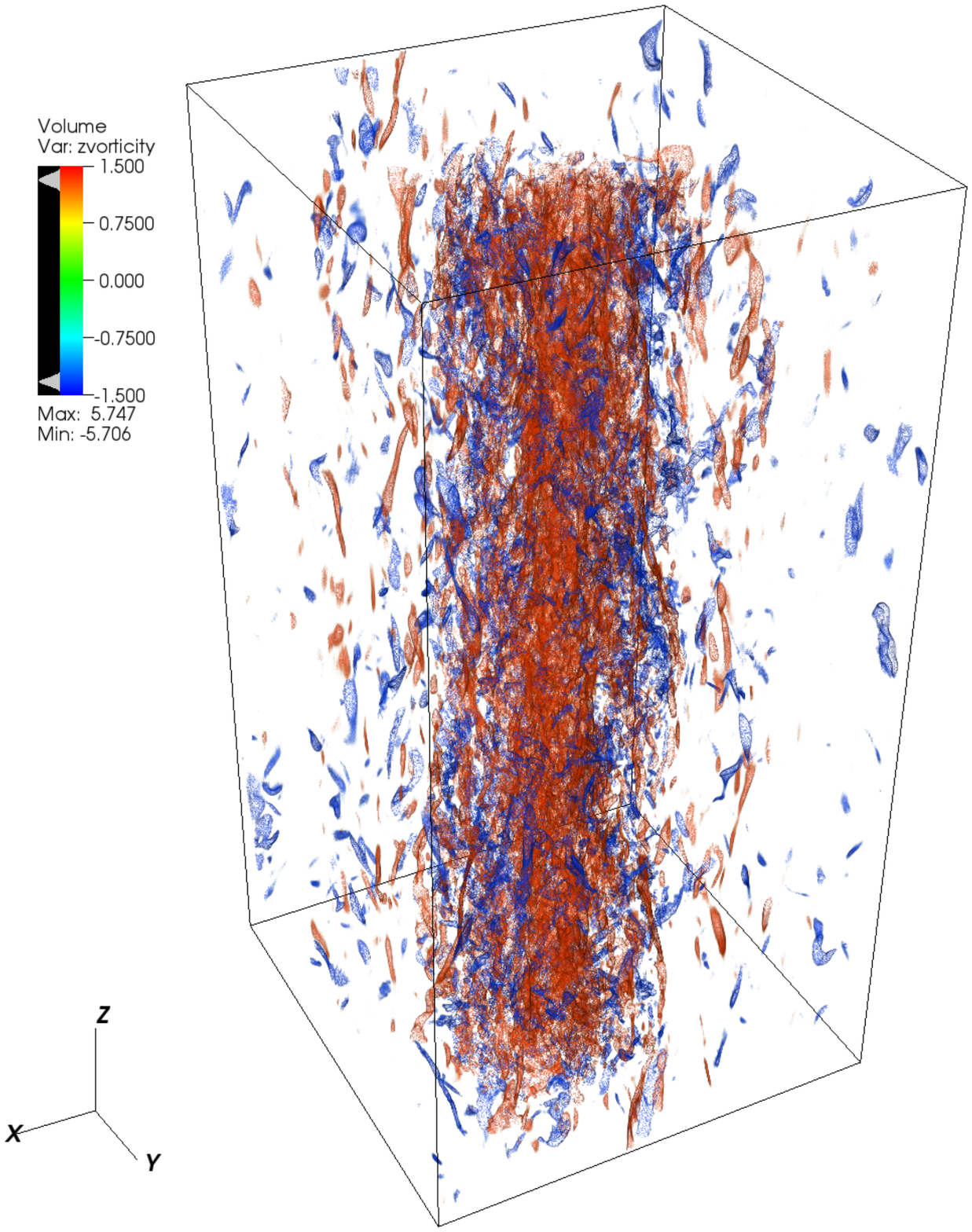}
\par\end{center}%
\end{minipage}\hfill{}%
\begin{minipage}[t]{0.45\columnwidth}%
\begin{center}
\includegraphics[scale=0.4]{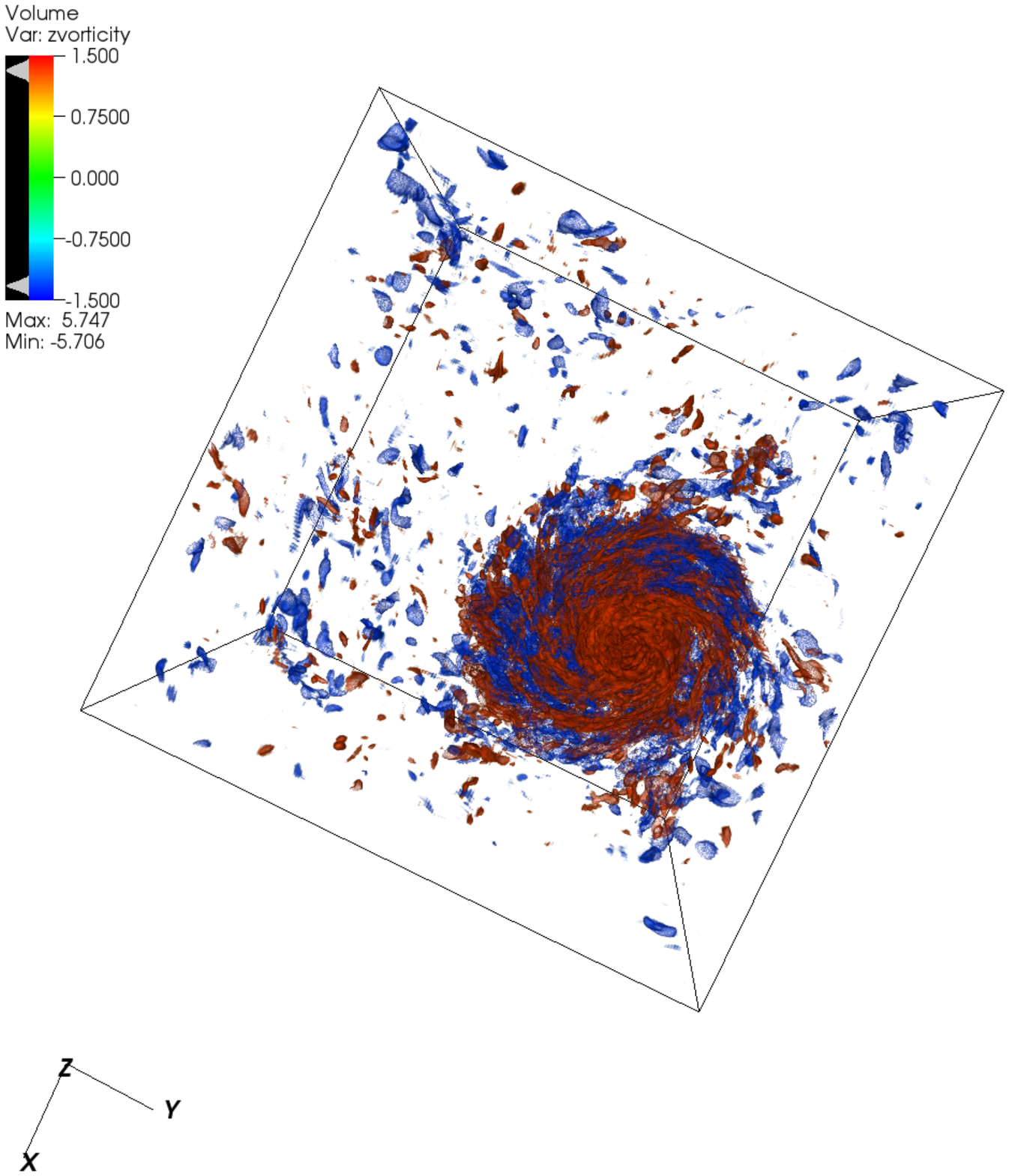}
\par\end{center}%
\end{minipage}
\par\end{raggedright}
\caption{\label{fig:vortexplots} (left) Snapshots of vertical vorticity $\omega_{z}=\left(\nabla\times\mathbf{u}\right)_{z}$;
(right) top view (along $x-y$ plane) - upper panels show two cyclonic
vortices at $t=217$ that later merge into a LSV shown in the bottom
two panels at $t=1300$.}
\end{figure}

\begin{figure}[H]
\begin{centering}
\includegraphics[scale=0.4]{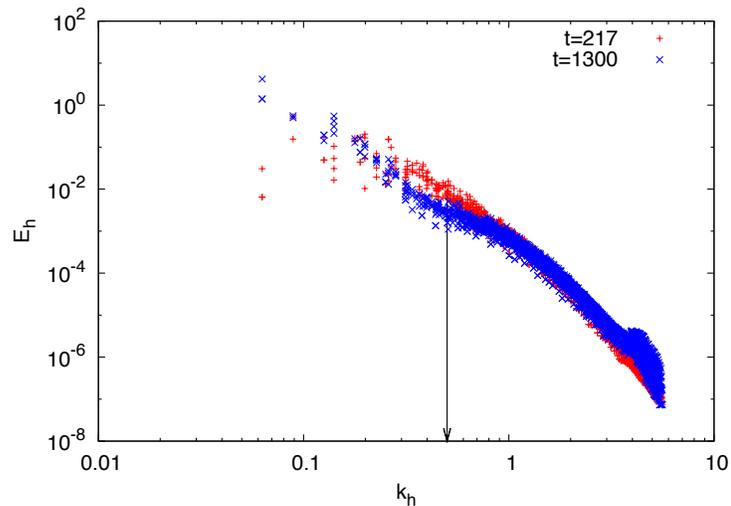}
\par\end{centering}
\caption{\label{fig:highres_enspec}Evolution of the horizontal energy, $E_{h}=\frac{1}{2}(u^{2}+v^{2})$,
for the lower resolution run (see text for details) at $R_{0}=1.45$
and ${\rm Ta}^{*}=10$, showing the gradual growth in energy of low
$k_{h}=\sqrt{l^{2}+m{}^{2}}$ modes with time, and the presence of
an inverse energy cascade; the arrow head shows the position of the
energy injection scale ($k_{h}\sim0.5$) corresponding to the typical
wavenumber of the fastest growing fingering modes.}
\end{figure}

While all the results reported so far were from high-resolution simulations,
we ran a few additional simulation at half their resolution for much
longer to see if any longer-term dynamics emerge. These runs are not
particularly under-resolved, so their dynamics are still reliable
i.e. the fingers and their structure are still well resolved. Interestingly,
one such run at a resolution of {\small{}$64\times64\times128$} for
$R_{0}=1.45$ and ${\rm Ta}^{*}=10$ shows a significant enhancement
in ${\rm Nu}{}_{\mu}$ over a long timescale ($\sim3000$ time units),
as shown in the left panel of Fig~\ref{fig:resolutioneffect}. 
It also shows a steady increase in the rms values of the vertical
velocity, chemical field ($\mu_{rms}$) as well as the vertical component
of the vorticity field, $\omega_{rms}$ (see right panel of Fig~\ref{fig:resolutioneffect}).
Fig~\ref{fig:Snapshots-of-chemical_uz} shows horizontal ($x-y$
plane) snapshots of the vertical velocity and the chemical fields
at time $t=1300$, and reveals the presence of a cyclonic large scale
vortex (hereafter, referred to as LSV). The LSV shows a substantial
enhancement in the concentration of high-$\mu$ fluid at its core,
associated with a strengthening of the downward vertical component
of velocity. It is to be noted that LSVs seen in other simulations
can also have the reverse situation, with low-$\mu$ material in their
core flowing upward. In both cases,
 this causes the enhancement in chemical transport measured through
the increase in ${\rm Nu}_{\mu}$ in Fig \ref{fig:resolutioneffect}.
Fig~\ref{fig:vortexplots} presents volume rendered snapshots of
the vertical vorticity in the flow at time $t=217$ and $t=1300$,
and clearly shows the emergence of long coherent cyclonic vortices
that later merge into a single cyclonic LSV spanning the entire height
and width of our domain. \\
In order to understand why this vortex forms and grows to fill the
domain, we inspect the horizontal energy spectrum of the simulation
(shown in left panel of Fig \ref{fig:highres_enspec}) which clearly
shows the development over time of a well-defined power law at low
horizontal wavenumber $k_{h}$, which is typical of an inverse energy
cascade associated with rotation. The inverse cascade draws its energy
at the injection scale $k_{h}\sim0.5$, which corresponds to the typical
wavenumber of the fastest-growing fingering modes. We can also estimate
the rate at which the LSV grows in strength by fitting an exponential
to the vorticity, $\omega_{rms}$ (between $t\sim500-2000)$ as
shown in the right panel of Fig \ref{fig:resolutioneffect}, which
gives a value of $\sim0.00039$ per unit time. The corresponding growth
timescale, which would be of order $3000$, is much larger than an
eddy turnover timescale (which is of order $10$), but much smaller
than the thermal or viscous diffusion timescales across the domain
(which are of order $10^{4}$ and $10^{5}$ respectively). 
The emergence of such LSVs is reminiscent of similar findings in the
work by \cite{2017ApJ...834...44M} (for the ODDC case) as well as
in studies of rapidly rotating convection (\cite{2014JFM...758..407G}).
In all such cases, LSVs are always seen to fill the domain, and have
also been interpreted as resulting from an inverse cascade (\cite{2018JFM...837R...4J}).

\section{Discussion\label{sec:Discussions}}

\subsection{Summary of our findings}

We have investigated the effect of rotation on the linear growth of
the fingering instability (Section 3) and found that rotation does
not affect the growth rate of the fastest-growing modes of the basic
linear instability. It does however influence its nonlinear evolution
and saturation. With the help of DNSs (Section 5) using the PADDI
code, we have measured the compositional transport rates of rotating
fingering convection in a parameter regime that approaches stellar
conditions. In general, we have found that rotation does not enhance
mixing by fingering convection contrary to our original expectations.
In fact, rotation seems to have a mild stabilizing effect on mixing.
The compositional transport rates predicted across a wide range of
rotation rates are consistently lower than the corresponding non-rotating
values measured in previous DNSs (\cite{2010ApJ...723..563D},\cite{2011JFM...677..530T,2013ApJ...768...34B}).
For simplicity, we restricted our present study to the polar configuration
only, so these findings need to be verified for the non-polar cases.
We suspect, however, that non-polar configurations will have even
weaker vertical mixing rates, simply by virtue of their geometry. 

We have observed a possible exception to this general finding for
a particularly turbulent (low $R_{0}$) and rapidly rotating (low
${\rm Ro}$) run in which coherent large scale structures naturally
emerge and gradually evolve to merge into a single cyclonic large
scale vortex spanning the entire computation domain. This LSV causes
a significant enhancement in the compositional transport rates by
concentrating high-$\mu$ material at its core that is advected downward.
Inspection of the horizontal kinetic energy spectrum demonstrates
that the LSV forms through a rotationally-driven inverse cascade that
draws its energy from the basic instability at the finger scale. The
LSV formation and dynamics are strongly reminiscent of those observed
in a variety of other rapidly rotating turbulent systems, such as
convection (\cite{2014JFM...758..407G,2018JFM...837R...4J}), stratified
turbulence (\cite{2013EL....10244006M,2017PhFl...29k1109O}) and oscillatory
double-diffusive convection (\cite{2017ApJ...834...44M}). 

\subsection{Implications for mixing in stars}

Our findings raise a tantalizing possibility: if these large-scale
vortices (LSVs) also form in the fingering regions of RGB stars, they
could substantially enhance the efficiency of mixing by fingering
convection, and thereby provide a self-consistent scenario to explain
the observed abundance changes on the upper RGB (\cite{2000A&A...354..169G,2007A&A...467L..15C}).
This raises the obvious question of whether LSVs would form under
more realistic stellar conditions. Studies of rapidly rotating convection
and oscillatory double-diffusive convection in the polar configuration
(i.e. with rotation aligned with gravity) have generally concluded
that LSVs are only observed in a rotationally constrained (low Rossby
number) and yet also strongly turbulent (high Reynolds number) regime
(\cite{2014JFM...758..407G,2018JFM...837R...4J,seshasayanan_alexakis_2018}),
which is also what we found here. The first of these conditions can
be understood by noting that strong rotation is required to trigger
an inverse energy cascade. However, rotation cannot be too strong
otherwise the flow becomes vertically invariant (through the Taylor-Proudman
constraint) and horizontal motions can only decay in that case. To
see this, note that the vertical component of the vorticity equation
(obtained by taking the curl of \ref{eq:momeq}) reduces to 
\begin{equation}
\frac{\partial\omega_{z}}{\partial t}+\mathbf{u}\cdot\nabla\omega_{z}={\rm Pr}\nabla^{2}\omega_{z}
\end{equation}
when motions are independent of $z$. In that limit, $\omega_{z}$
must ultimately decay with time (since this advection-diffusion equation
contains no source term), which in turn shows that horizontal motions
must necessarily decay as well. In other words, the flow must remain
sufficiently three-dimensional to continually feed energy into the
inverse cascade and maintain the vortex against viscous decay, hence
the need for a sufficiently large Reynolds number. 

In Section 4 (combined with the results of Figure \ref{fig:Comparison-of-Rossby}),
we showed that the fingering regions of RGB stars would indeed satisfy
the low Rossby number requirement, with estimated values in the range
$10^{-3}-1$ for rapid and moderate rotators. To estimate the Reynolds
number ${\rm Re}$ expected in these regions, we use a similar argument
as in Section 4. Since ${\rm Re}=UL/\nu$, where $U$ and $L$ are
the characteristic velocities and lengthscale of fingering flows (given
by equation \ref{eq:Brownetalmodel}) and $\nu$ is the viscosity,
then 
\begin{equation}
{\rm Re\simeq\sqrt{\frac{\tau}{r{\rm Pr}}}}\simeq\frac{1}{\sqrt{{\rm Pr}(R_{0}-1)}}.
\end{equation}
According to this estimate, using ${\rm Pr}\sim10^{-6}$ and $R_{0}\sim10^{3}$,
as before, we find that ${\rm Re}\sim10^{2}$, which should indeed
be sufficiently high for LSVs to form. 

We therefore conclude that fingering regions of RGB stars can indeed
potentially be the home of large-scale vortices near the poles, which
would cause a very substantial enhancement of the compositional fluxes
and could in turn explain the observed evolution of the surface abundances
after the luminosity bump. 

Of course, much remains to be done to confirm this scenario. In particular,
recent results on the formation of large-scale vortices in other systems
such as rotating convection and oscillatory double-diffusive convection
suggest that they may not develop (1) at lower latitudes (\cite{2017ApJ...834...44M}),
and (2) unless the computational domain has a unit aspect ratio (\cite{2018JFM...837R...4J}).
In these cases, large-scale horizontal jets form instead. Whether
these would also be more common in the case of rotating fingering
convection remains to be determined, but is likely. Whether compositional
transport would similarly be enhanced in the presence of jets or not
also remains to be determined, but also seems likely. These questions
will be answered in future work, as they require substantial computational
resources to fully explore.

\section{Conclusions\label{sec:conclusion}}

The simulations presented here clearly point out the need to understand
better the interplay of different physical mechanisms in order to
provide robust estimates of mixing to be used in stellar evolutionary
calculations. Most modern stellar evolution codes treat mixing processes
independently, by computing a simple diffusion coefficient for each
one of them and then adding them together (\cite{2010A&A...521A...9C,2011A&A...536A..28L,2017A&A...606A..55M,2018ApJS..234...34P}).
This study reveals that although rotation and fingering convection
can indeed be fairly well understood independently in some regimes,
other regimes exist in which they strongly reinforce one another.
We showed that this regime is precisely the one that is relevant for
the RGB ``extra-mixing'' problem. If indeed LSVs form in the radiative
zone above the H-burning shell in the interiors of RGB stars, they
could greatly enhance transport and provide a self-consistent scenario
explaining the observed abundances changes on the upper RGB which
non-rotating model predictions fail to do. This is the only possible
scenario in the context of the ``missing-mixing'' problem of the
RGBs which could work to explain the observed change in abundances
of red-giants above the luminosity bump in a self-consistent way without
the need to invoke physical mechanisms that are not specific to this
particular evolutionary phase (\cite{2007A&A...476L..29C,2009ApJ...696.1823D}).We
aim to explore the emergence of these LSVs across a wider range of
parameter space in a future work (Sengupta \& Garaud, in preparation)
to make more systematic predictions for the conditions in which one
can expect them to form.

\acknowledgments S. S. and P. G. were funded from NSF AST 1412951.
We thank S. Stellmach for the use of the PADDI code. The simulations
were performed on the Hyades supercomputer, purchased using an NSF
MRI grant. Figures \ref{fig:Snapshots-of-verticalvel} and \ref{fig:vortexplots}
were rendered using VisIt, a product of the Lawrence Livermore National
Laboratory.
\begin{singlespace}

\end{singlespace}

\section*{Appendix}

\section*{Effect of domain size}

\begin{figure}
\begin{centering}
\includegraphics[width=0.6\textwidth]{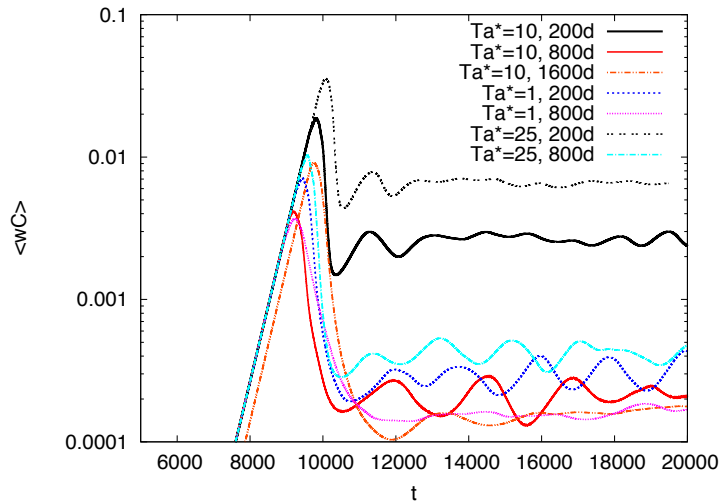}
\par\end{centering}
\caption{\label{fig:Compositional-fluxes-highR0}Compositional fluxes for simulations
using domain heights of $200d$ and $800d$ at $R_{0}=9.1$ for ${\rm Ta}^{*}=1.0,10.0$}

\end{figure}

From Table~1, we can note that for $R_{0}=5$, the difference in
the compositional Nusselt numbers between two simulations for which
the height of the domain differs by factor of $2,$ is at most few
percent even for our highest ${\rm Ta}^{*}$ runs. However, for a
more extreme choice of $R_{0}=9.1$, Fig \ref{fig:Compositional-fluxes-highR0}
shows that using our default domain size at ${\rm Ta}^{*}=10$ or
$25$ gives compositional fluxes that differ by up to an order of
magnitude from those obtained by using a taller domain ($100d\times100d\times800d$).
A similar effect was also observed by \cite{2011PhDT........23T}
even for the non-rotating case for very high values of $R_{0}$ close
to the marginal stability threshold of $\frac{1}{\tau}$.\\
\begin{figure}
\begin{raggedright}
\includegraphics[width=1\textwidth]{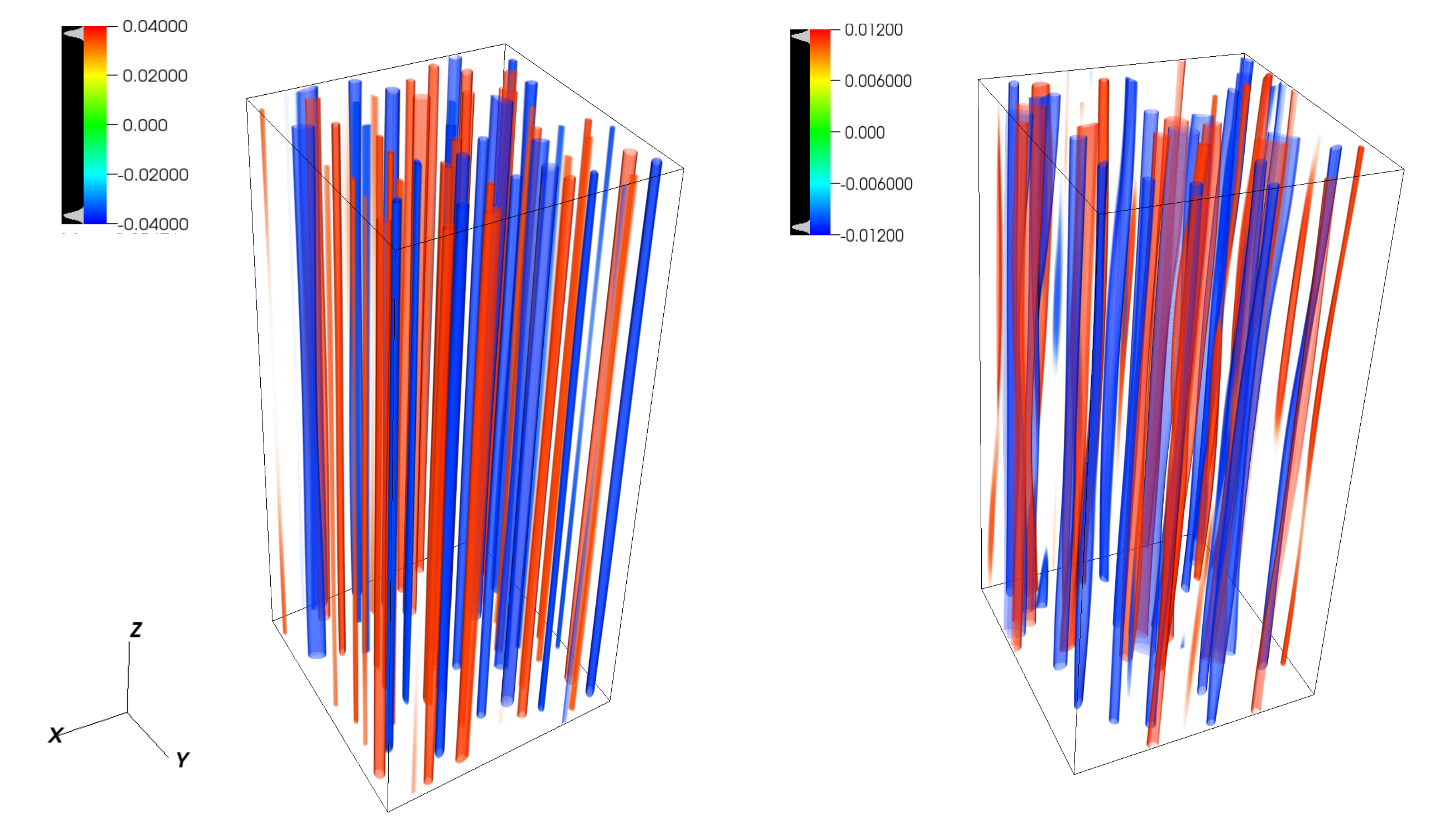}
\par\end{raggedright}
\caption{\label{fig:domainheight} Vertical velocity fields (after saturation)
for $R_{0}=9.1$ and ${\rm Ta}^{*}=10$ for domain heights of $200d$(left)
and $800d$ (right), showing the need for using an elongated domain
(in the vertical direction) at high $R_{0}$.}
\end{figure}

Fig~\ref{fig:domainheight} shows snapshots of the vertical velocity
in two simulations for $R_{0}=9.1$, ${\rm Ta}^{*}=10$ - the left
panel using our default domain size and the right panel with a $100d\times100d\times800d$
domain. The $200d$-tall domain has fingers that are perfectly vertical,
whereas the $800d$-tall domain\footnote{The image for the $800d$ tall domain has been compressed vertically
by a factor of 4 to show it on the same scale as the $200d$ tall
domain.}, shows fingers that no longer remain perfectly vertical. We conjecture
that the fastest growing wavelength of the shear instability between
upflowing and downflowing fingers increases with increasing rotation
rate. When the latter exceeds the domain size, the shear instability
is suppressed and the transport is vastly enhanced. This effect is
artificial, however, and must be avoided by making sure the domain
is indeed tall enough to contain the shear-unstable modes.
\end{document}